\documentclass[aps,prl,twocolumn,superscriptaddress,longbibliography]{revtex4-1}

\usepackage{graphicx}
\usepackage{dcolumn}
\usepackage{bm}
\usepackage{amssymb}
\usepackage{microtype}
\usepackage{xfrac}
\usepackage{array}

\newcommand{\ket}[1]{\ensuremath{|#1\rangle}}

\begin{document}

\title{Dipolar-Octupolar Ising Antiferromagnetism in Sm$_2$Ti$_2$O$_7$:\\ A Moment Fragmentation Candidate}

\author{C. Mauws}
\affiliation{Department of Chemistry, University of Manitoba, Winnipeg R3T 2N2, Canada}
\affiliation{Department of Chemistry, University of Winnipeg, Winnipeg R3B 2E9, Canada}

\author{A.~M.~Hallas}
\affiliation{Department of Physics and Astronomy, McMaster University, Hamilton L8S 4M1, Canada}
\affiliation{Department of Physics and Astronomy and Rice Center for Quantum Materials, Rice University, Houston, TX, 77005 USA}

\author{G.~Sala}
\affiliation{Neutron Scattering Division, Oak Ridge National Laboratory, Oak Ridge, Tennessee 37831, USA}

\author{A.~A.~Aczel}
\affiliation{Neutron Scattering Division, Oak Ridge National Laboratory, Oak Ridge, Tennessee 37831, USA}

\author{P.~M.~Sarte}
\affiliation{School of Chemistry, University of Edinburgh, Edinburgh EH9 3FJ, United Kingdom}
\affiliation{Centre for Science at Extreme Conditions, University of Edinburgh, Edinburgh EH9 3FD, United Kingdom}

\author{J.~Gaudet}
\affiliation{Department of Physics and Astronomy, McMaster University, Hamilton L8S 4M1, Canada}

\author{D.~Ziat}
\affiliation{Institut Quantique and D\'{e}partement de Physique, Universit\'{e} de Sherbrooke, Sherbrooke, Qu\'{e}bec J1K 2R1, Canada}

\author{J.~A.~Quilliam}
\affiliation{Institut Quantique and D\'{e}partement de Physique, Universit\'{e} de Sherbrooke, Sherbrooke, Qu\'{e}bec J1K 2R1, Canada}

\author{J.~A.~Lussier}
\affiliation{Department of Chemistry, University of Manitoba, Winnipeg R3T 2N2, Canada}

\author{M. Bieringer}
\affiliation{Department of Chemistry, University of Manitoba, Winnipeg R3T 2N2, Canada}

\author{H.~D. Zhou}
\affiliation{Department of Physics and Astronomy, University of Tennessee-Knoxville, Knoxville 37996-1220, United States}
\affiliation{National High Magnetic Field Laboratory, Florida State University, Tallahassee 32306-4005, United States}

\author{A.~Wildes}
\affiliation {Institut Laue-Langevin, 71 avenue des Martyrs, CS 20156, 38042 Grenoble Cedex 9, France}

\author{M.~B.~Stone}
\affiliation{Neutron Scattering Division, Oak Ridge National Laboratory, Oak Ridge, Tennessee 37831, USA}

\author{D.~Abernathy}
\affiliation{Neutron Scattering Division, Oak Ridge National Laboratory, Oak Ridge, Tennessee 37831, USA}

\author{G.~M.~Luke}
\affiliation{Department of Physics and Astronomy, McMaster University, Hamilton L8S 4M1, Canada}
\affiliation{Canadian Institute for Advanced Research, Toronto M5G 1M1, Canada}
\affiliation{TRIUMF, 4004 Wesbrook Mall, Vancouver, British Columbia, Canada V6T 2A3}

\author{B.~D.~Gaulin}
\affiliation{Department of Physics and Astronomy, McMaster University, Hamilton L8S 4M1, Canada}
\affiliation{Canadian Institute for Advanced Research, Toronto M5G 1M1, Canada}

\author{C.~R.~Wiebe}
\affiliation{Department of Chemistry, University of Manitoba, Winnipeg R3T 2N2, Canada}
\affiliation{Department of Chemistry, University of Winnipeg, Winnipeg R3B 2E9, Canada}
\affiliation{Department of Physics and Astronomy, McMaster University, Hamilton L8S 4M1, Canada}
\affiliation{Canadian Institute for Advanced Research, Toronto M5G 1M1, Canada}

\date{\today}

\begin{abstract}
Over the past two decades, the magnetic ground states of all rare earth titanate pyrochlores have been extensively studied, with the exception of Sm$_2$Ti$_2$O$_7$. This is, in large part, due to the very high absorption cross-section of naturally-occurring samarium, which renders neutron scattering infeasible. To combat this, we have grown a large, isotopically-enriched single crystal of Sm$_2$Ti$_2$O$_7$. Using inelastic neutron scattering, we determine that the crystal field ground state for Sm$^{3+}$ is a dipolar-octupolar doublet with Ising anisotropy. Neutron diffraction experiments reveal that Sm$_2$Ti$_2$O$_7$ orders into the all-in, all-out magnetic structure with an ordered moment of 0.44(7) $\mu_B$ below $T_N=0.35$~K, consistent with expectations for antiferromagnetically-coupled Ising spins on the pyrochlore lattice. Zero-field muon spin relaxation measurements reveal an absence of spontaneous oscillations and persistent spin fluctuations down to 0.03~K. The combination of the dipolar-octupolar nature of the Sm$^{3+}$ moment, the all-in, all-out ordered state, and the low-temperature persistent spin dynamics make this material an intriguing candidate for moment fragmentation physics. 
\end{abstract}

\maketitle



\begin{figure}[tbp]
\linespread{1}
\par
\includegraphics[width=3.3in]{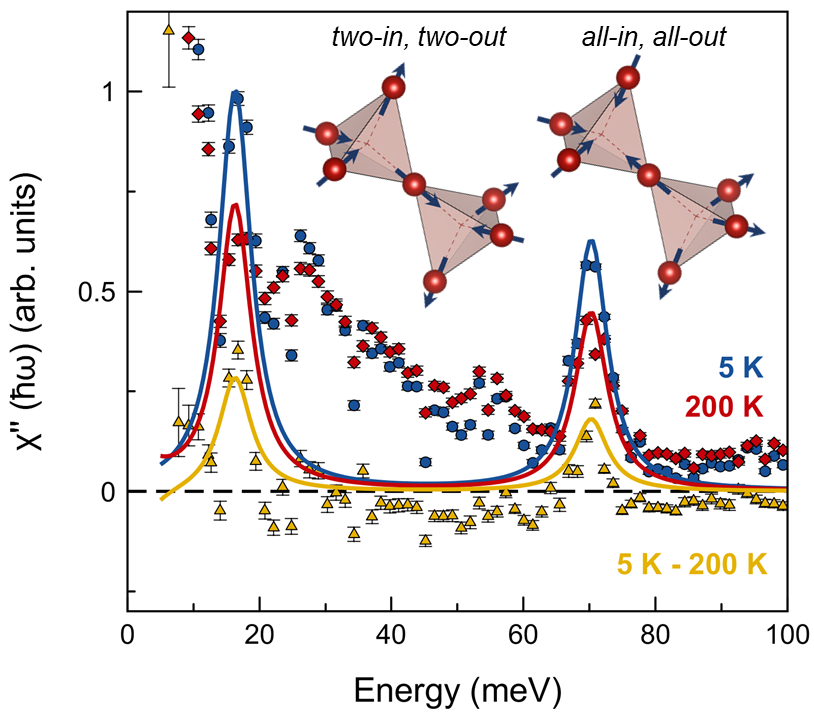}
\par
\caption{Inelastic neutron scattering measurements of the crystal electric field (CEF) excitations in Sm$_2$Ti$_2$O$_7$ at 5~K (red) and 200~K (blue). The temperature difference (yellow) confirms the presence of two CEF levels, at 16.3(5)~meV and 70.0(5)~meV. The fits to the data with our CEF model are given by the solid lines and reveal the Ising nature of the Sm$^{3+}$ moments in Sm$_2$Ti$_2$O$_7$. The insets show the ferromagnetic Ising spin configuration (two-in, two-out) and the antiferromagnetic Ising spin configuration (all-in, all-out).}
\label{CEF}
\end{figure}

Rare earth titanate pyrochlores of the form $R_2$Ti$_2$O$_7$ have long been a centerpiece in the study of geometrically-frustrated magnetism~\cite{gardner2010magnetic}. In this family of materials, the magnetism is carried by the $R^{3+}$ rare earth ions, which decorate a network of corner-sharing tetrahedra.
The study of this family has led to the discovery of a range of fascinating ground states such as the dipolar spin ice state, which was first observed in Ho$_2$Ti$_2$O$_7$ and Dy$_2$Ti$_2$O$_7$~\cite{harris1997geometrical,ramirez1999zero,bramwell2001spin}. Here local Ising anisotropy combines with dominant dipolar interactions, which are ferromagnetic at the nearest neighbour level on the pyrochlore lattice
~\cite{den2000dipolar}. The spin ice state is characterized by individual tetrahedra obeying two-in, two-out ``ice rules", wherein two spins point directly towards the tetrahedron's center and the other two spins point outwards (left inset of Fig.~\ref{CEF}). This configuration can be achieved in six equivalent ways for a single tetrahedron, giving rise to a macroscopic degeneracy for the lattice as a whole. In other titanates, where the rare earth moments are smaller than in Ho$_2$Ti$_2$O$_7$ and Dy$_2$Ti$_2$O$_7$, dipolar interactions become less important and exchange interactions tend to dominate. This is exactly the case when $R =$~Sm$^{3+}$ ($\sim1~\mu_B$), where the magnetic moment is reduced by a factor of ten from $R =$~Ho$^{3+}$ and Dy$^{3+}$ ($\sim10~\mu_B$), corresponding to dipolar interactions that are weaker by two orders of magnitude.

In this letter we show that anitferromagnetically coupled Ising spins with negligible dipolar interactions give rise to an all-in, all-out (AIAO) magnetic ground state in Sm$_2$Ti$_2$O$_7$. The AIAO structure is characterized by adjacent tetrahedra alternating between all spins pointing inwards and all spins pointing outwards (right inset of Fig.~\ref{CEF}). Unlike the ferromagnetic spin ice state, the antiferromagnetic AIAO state does not give rise to a macroscopic degeneracy; placing a single spin as ``in'' or ``out'' is enough to uniquely constrain the orientations of all other spins on the lattice. A host of neodymium pyrochlores with varying non-magnetic $B$ sites also display the AIAO ground state, Nd$_2B_2$O$_7$ ($B=$~Sn, Zr, Hf) ~\cite{bertin2015nd,xu2015magnetic,lhotel2015fluctuations,anand2015observation}. Nd$_2$Zr$_2$O$_7$ is a particularly interesting case as magnetic Bragg peaks from the AIAO structure and disordered, spin ice-like diffuse scattering coexist at low temperatures~\cite{petit2016observation}. This exotic phenomenology 
has been termed moment fragmentation~\cite{brooks2014magnetic}. Recent theoretical work~\cite{benton2016quantum} has argued that the origin of this effect is the peculiar dipolar-octupolar symmetry of the Nd$^{3+}$ ground state doublet~\cite{xu2015magnetic,lhotel2015fluctuations}. When combined with an AIAO ground state, the symmetry properties of this dipolar-octupolar doublet allow the decoupling of the divergence-full (AIAO) and divergence-free (spin ice) fluctuations~\cite{benton2016quantum}. Here we use neutron spectroscopy to determine the dipolar-octupolar nature of the crystal field ground state doublet of Sm$_2$Ti$_2$O$_7$ and use neutron diffraction to show that it orders into an AIAO structure below $T_N=0.35$~K. Muon spin relaxation measurements reveal persistent spin dynamics within the magnetically ordered state, down to 0.03~K. Thus, we demonstrate that Sm$_2$Ti$_2$O$_7$ possesses the requisite ingredients for moment fragmentation physics.


In contrast to the extensive studies that have been performed on the other magnetic titanate pyrochlores, $R_2$Ti$_2$O$_7$ ($R=$~Gd, Tb, Dy, Ho, Er, Yb), the magnetic properties of Sm$_2$Ti$_2$O$_7$ have remained largely unexplored. Prior studies of Sm$_2$Ti$_2$O$_7$ were limited to bulk property measurements in the paramagnetic regime, above 0.5~K, which revealed weak antiferromagnetic interactions ($\theta_{CW} = -0.26$~K)~\cite{singh2008manifestation}. While the other above-mentioned titanate pyrochlores have been the subjects of a plethora of elastic and inelastic neutron scattering experiments, equivalent experiments on Sm$_2$Ti$_2$O$_7$ are daunting. The first reason is the size of the Sm$^{3+}$ magnetic moment; the Lande $g$-factor associated with the $4f^5$ electronic configuration is its smallest possible non-zero value ($g_J = \sfrac{2}{7}$), giving rise to small moments even in the absence of crystal field effects (which make the moment smaller still). This small magnetic moment represents a significant hindrance because scattered neutron intensity varies as the moment squared. Compounding this effect is that naturally-occurring samarium is a very strong neutron absorber due to the presence of $^{149}$Sm at the 13.9\% level ($\sigma_{abs} = 42,000$ barns). Neutron scattering measurements of the type we report here are only possible with a sample isotopically-enriched with $^{154}$Sm ($\sigma_{abs} = 8.4$ barns). However, the neutron absorption cross section of $^{149}$Sm is so high that even trace amounts result in a sample that is still strongly absorbing by neutron scattering standards. 


We grew a large single crystal of Sm$_2$Ti$_2$O$_7$ with the optical floating zone technique using 99.8\% enriched $^{154}$Sm$_2$O$_3$ (Cambridge Isotopes). Low-temperature heat capacity measurements were performed using the quasi-adiabatic technique. Neutron diffraction measurements were performed on the D7 polarized diffuse scattering spectrometer at the Institute Laue-Langevin and beam line HB-1A at the High Flux Isotope Reactor at Oak Ridge National Laboratory (ORNL). Inelastic neutron scattering measurements were performed on the ARCS~\cite{abernathy2012design} and SEQUOIA~\cite{granroth2010sequoia} spectrometers at the Spallation Neutron Source at ORNL. Muon spin relaxation measurements were carried out at TRIUMF. Further experimental details are provided in the Supplementary Materials. 


\begin{table}
\caption{Result of the CEF analysis for Sm$_2$Ti$_2$O$_7$, calculated within a point charge model and then refined by fitting the two experimentally observed CEF excitations.}
\begin{ruledtabular}
\begin{tabular}{c|cccc}
E$_{obs}$ (meV) & E$_{fit}$ (meV) & $\ket{\pm\sfrac{5}{2}}$&$\ket{\pm\sfrac{3}{2}}$&$\ket{\pm\sfrac{1}{2}}$ \\ \hline
0.0 & 0.0 & 0 & 1 & 0 \\
16.3(5) & 16.5 & 0 & 0 & 1 \\
70.0(5) & 70.3 & 1& 0 & 0\\
\end{tabular}
\end{ruledtabular}
\label{tab: 1} 
\end{table}


The Hund's rules ground state for Sm$^{3+}$ is $J = \sfrac{5}{2}$. Accordingly, in the reduced symmetry environment of the pyrochlore lattice, the $2J+1 = 6$ states split into three Kramers' doublets, one of which forms the crystal electric field (CEF) ground state. Inelastic neutron scattering (INS) measurements on Sm$_2$Ti$_2$O$_7$, which are presented in Fig.~\ref{CEF} and Fig. S2, show intense excitations at 16.3(5) and 70.0(5) meV corresponding to transitions to the excited CEF doublets. The lower energy excitation is consistent with one of the modes previously identified in Raman scattering experiments by Singh \emph{et al}~\cite{singh2008manifestation}. However, other modes observed in Raman scattering and originally attributed to additional CEF excitations are not visible in our INS data. Malkin {\it et al.} attempted to determine the crystal field parameters of Sm$_2$Ti$_2$O$_7$ by modeling magnetic susceptibility data~\cite{malkin2010static}. This work predicts CEF levels at 21.4 and 26.4 meV, both of which are inconsistent with our INS data. It is worth noting that Sm$^{3+}$ has a rather atypical form factor, which rather than monotonically decreasing with $Q$ instead reaches its maximum value near 5~\AA$^{-1}$. Both of the CEF transitions observed here obey this form factor (see Supplmentary Materials). These two excitated states account for the full manifold associated with the $J=\sfrac{5}{2}$ ground state multiplet.

Next, we modeled the INS data in order to extract the crystal field Hamiltonian. This analysis is complicated by the strong residual absorption of $^{149}$Sm in the isotopically-enriched single crystal. This issue was addressed by performing an absorption correction with Monte Carlo ray tracing simulations using MCViNE~\cite{lin2016mcvine}. In the case of Sm$^{3+}$, the Hund's rules $J$ manifold is separated from the first excited spin-orbit manifold by $\lambda(J+1)\approx500$~meV~\cite{blume1964theory}. Incorporating this higher manifold into our analysis would require the introduction of four additional free parameters. This would result in an under constrained parameterization of the CEF Hamiltonian and thus, we have neglected it here. Further details of these calculations and the subsequent determination of the CEF eigenvalues and eigenvectors are presented in the Supplementary Material. 

The CEF parameters that provide the best fit to our INS data for Sm$_2$Ti$_2$O$_7$ within a point charge approximation are: $B_{20} = 3.397$~meV, $B_{40} = 0.123$~meV, and $B_{43} = 8.28 \cdot 10^{-8}$~meV. Table~\ref{tab: 1} shows the resulting CEF eigenvectors and eigenvalues. Our refinement gives a ground-state doublet of pure $\ket{m_J = \pm \sfrac{3}{2}}$ character. The three-fold rotational symmetry at the rare earth site implies that states within a time-reversal symmetry-paired Kramers doublet must be composed of $m_J$ basis states separated by three units. Accordingly, in our case where the maximum $m_J=\sfrac{5}{2}$, it follows that the doublet composed of $\ket{m_J = \pm \sfrac{3}{2}}$ cannot be coupled to any other basis state and is hence, necessarily pure. The symmetry nature of this doublet imparts it with an exotic character: while two components of the pseudospin transform like a magnetic dipole, the third component transforms as a component of the magnetic octupole tensor~\cite{huang2014quantum}. Thus, the ground state doublet in Sm$_2$Ti$_2$O$_7$ is termed a dipolar-octupolar doublet. This result distinguishes Sm$_2$Ti$_2$O$_7$ from other antiferromagnetic Kramers $R_2$Ti$_2$O$_7$ pyrochlores ($R=$~Er~\cite{gaudet2017effect} and Yb~\cite{gaudet2015neutron}), which possess ground state doublets that transform simply as a magnetic dipole, effectively mimicking a true $S=\sfrac{1}{2}$. Our refined $g$-tensor gives $g_{z} =0.857(9)$ and $g_{xy} = 0.0$, corresponding to Ising anisotropy, where the spins point along their local [111] direction, which connects the vertices of the tetrahedron to its center (inset of Fig.~\ref{CEF}). The magnetic moment within the ground state doublet of Sm$^{3+}$ is $\mu_{\text{CEF}} = 0.43(6)~\mu_B$. 


\begin{figure}[tbp]
\linespread{1}
\par
\includegraphics[width=3.3in]{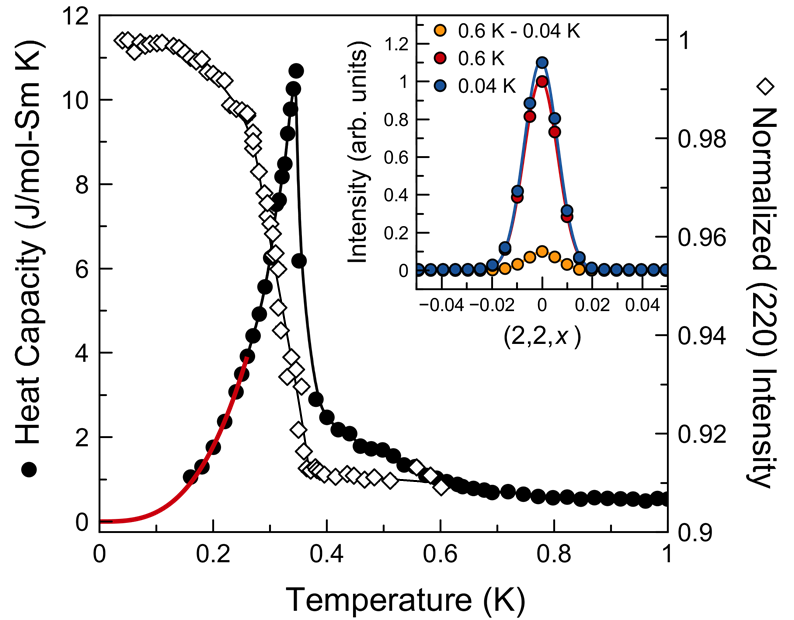}
\par
\caption{Sm$_2$Ti$_2$O$_7$ undergoes a long-range magnetic ordering transition at $T_N = 0.35$~K. The black circles represent the heat capacity, which shows a sharp anomaly at $T_N$ and a $T^3$ dependence at the lowest temperatures, as indicated by the red line. The open diamonds represent the intensity of the (220) Bragg peak, which shows an abrupt increase at $T_N$. The black lines are guides to the eye. The inset shows a scan over the (220) Bragg peak above and below $T_N$, where the enhanced intensity corresponds to the formation of a magnetic Bragg peak.}
\label{HeatCapacity}
\end{figure}

Finally, as originally discussed in Ref.~\cite{Hutchings1964227}, we can take advantage of the fact that extensive CEF studies have been performed on other rare earth titanate pyrochlores \cite{rosenkranz2000crystal, bertin2012crystal, ruminy2016crystal, gaudet2015neutron, gaudet2017effect, princep2015crystal}, allowing us to use scaling arguments. An especially good starting point is Er$^{3+}$ in Er$_2$Ti$_2$O$_7$, which has a large total angular momentum, $J=15/2$. This material has seven excited crystal field levels, all of which were observed in a recent INS study, leading to a highly constrained CEF Hamiltonian~\cite{gaudet2017effect}. Armed with these results, scaling arguments give us qualitatively good agreement with the known CEF manifolds for $R_2$Ti$_2$O$_7$ ($R=$~Ho, Tb, and Yb)~\cite{gaudet2017effect}. When applied to Sm$_2$Ti$_2$O$_7$, these same scaling arguments predict the CEF ground state to be pure $\ket{m_J = \pm \sfrac{3}{2}}$ with a large energy gap to the first excited state, consistent with our experimental determination.



We next turn to the low-temperature collective magnetic properties of Sm$_2$Ti$_2$O$_7$. The heat capacity of Sm$_2$Ti$_2$O$_7$, shown in Fig.~\ref{HeatCapacity}(a), contains a lambda-like anomaly at $T_N = 0.35$~K, indicative of a second-order phase transition to a long-range magnetically ordered state. This ordering transition was not observed in previous studies as their characterization measurements did not extend below 0.5~K~\cite{singh2008manifestation}. The low temperature region of the anomaly, below 0.3~K, is well-fit by a $T^3$ power law, consistent with gapless, three-dimensional antiferromagnetic spin waves. In order to compute the entropy release associated with this anomaly, we extrapolate the $T^3$ behavior to 0~K. Then, an integration of $C/T$ up to 1 K returns an entropy of $0.84 \cdot R\ln{2}$, close to the full $R\ln{2}$ expected for a well-isolated Kramers doublet. Thus, a small fraction of the entropy release in this system may be taking place at temperatures above 1~K or some fraction of the moment may remain dynamic below $T_N$.



\begin{figure}[tbp]
\linespread{1}
\par
\includegraphics[width=3.3in]{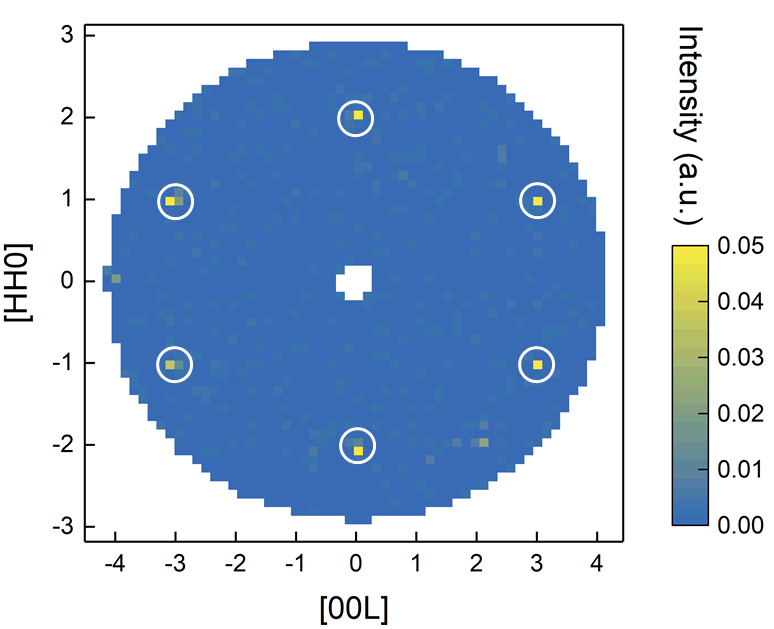}
\par
\caption{Spin flip channel of polarized neutron scattering measurements on Sm$_2$Ti$_2$O$_7$ in the (HHL) scattering plane at 0.05~K. Magnetic Bragg peaks are observed to form on the (220) and (113) positions and are absent at the (111), (002), (222) and (004) positions. Symmetry analysis of this magnetic diffraction pattern reveals that Sm$_2$Ti$_2$O$_7$ is ordering into the $\Gamma_3$ AIAO state with an ordered moment of $\mu_{ord}=0.44(7)~\mu_B$. Note that weak bleed-through of nuclear Bragg intensity has been corrected by subtracting a high temperature (4 K) data set.}
\label{SpinFlip}
\end{figure}

We used the D7 polarized neutron scattering spectrometer at the ILL to search for magnetic diffuse scattering in Sm$_2$Ti$_2$O$_7$. While none could be resolved above or below $T_N$, we did observe the formation of magnetic Bragg peaks at the (220) and (113) positions in the spin flip channel (Figure~\ref{SpinFlip}). The observed magnetic Bragg reflections were indexed against the possible $\vec{k}=0$ ordered structures for the $16c$ Wyckoff position in the $Fd\bar{3}m$ pyrochlore lattice (Table~\ref{tab: 2}). The errors on the observed peak intensities are rather high due to the small magnetic signals ($\mu_{\text{ord}} \leq \mu_{\text{CEF}} = 0.43~\mu_B$) located on large nuclear Bragg peaks, the absorption from residual $^{149}$Sm, as well as the relatively poor $Q$-resolution of a diffuse scattering instrument. However, as can be seen by careful examination of Table~\ref{tab: 2}, the observed magnetic Bragg reflections nicely map onto the $\Gamma_3$ irreducible representation. All other representations can be ruled out by the absence of magnetic reflections at the (002) and (111) positions in the experimental data. $\Gamma_3$ corresponds to the AIAO magnetic structure (right inset of Fig~\ref{CEF}), which is the expected result when Ising anisotropy is combined with net antiferromagnetic exchange interactions. The neutron order parameter, shown in Fig.~\ref{HeatCapacity}, reveals a sharp onset below $T_N = 0.35$~K, fully-consistent with the anomaly observed in the heat capacity.

\begin{table}[tbp]
\caption{Bragg peak intensities for the possible $\vec{k}=0$ magnetic structures for Sm$_2$Ti$_2$O$_7$. The best agreement is obtained with the $\Gamma_3$ all-in all-out structure. }
\label{tab: 2}
\begin{ruledtabular}
\begin{tabular}{ccccccc}
 &(111)&(002)&(222)&(220)&(113)&(004)\\ \hline
 Observed&0&0&0&1.0$\pm$0.4&0.78$\pm$0.27&0 \\ \hline
 $\Gamma_3$ &0&0&0&1.00&0.66&0\\
 $\Gamma_5$ &0.88&0&0&1&0.35&0\\
 $\Gamma_7$ &0.52&1.00&0.44&0.11&0&0\\
 $\Gamma_9$ [110] &0.69&1.00&0.44&0.43&0.51&0.67\\
 $\Gamma_9$ [100] &0.06&0.37&0.16&0.44&0.76&1.00\\
\end{tabular}
\end{ruledtabular}
\end{table}

While the D7 data allowed a definitive determination of the magnetic structure of Sm$_2$Ti$_2$O$_7$, it is not appropriate for estimating the value of the ordered moment due to the coarse $Q$-resolution of the instrument. The triple axis spectrometer HB-1A, with its significantly-improved $Q$-resolution, was therefore used for this purpose. Since HB-1A uses an unpolarized neutron beam, magnetic intensity was only observed at the (220) Bragg peak position in this experiment, which corresponds to the strongest magnetic reflection expected for the AIAO magnetic structure but also a relatively weak nuclear Bragg peak. We determined the Sm$^{3+}$ ordered magnetic moment by comparing the ratio of the magnetic intensity to the nuclear intensity at this Bragg position. This procedure, which incorporated both the $j_0$ and $j_2$ spherical Bessel function contributions to the Sm$^{3+}$ magnetic form factor, yielded an ordered moment of $\mu_{\text{ord}} = 0.44(7)~\mu_B$.



\begin{figure}[tbp]
\linespread{1}
\par
\includegraphics[width=3.3in]{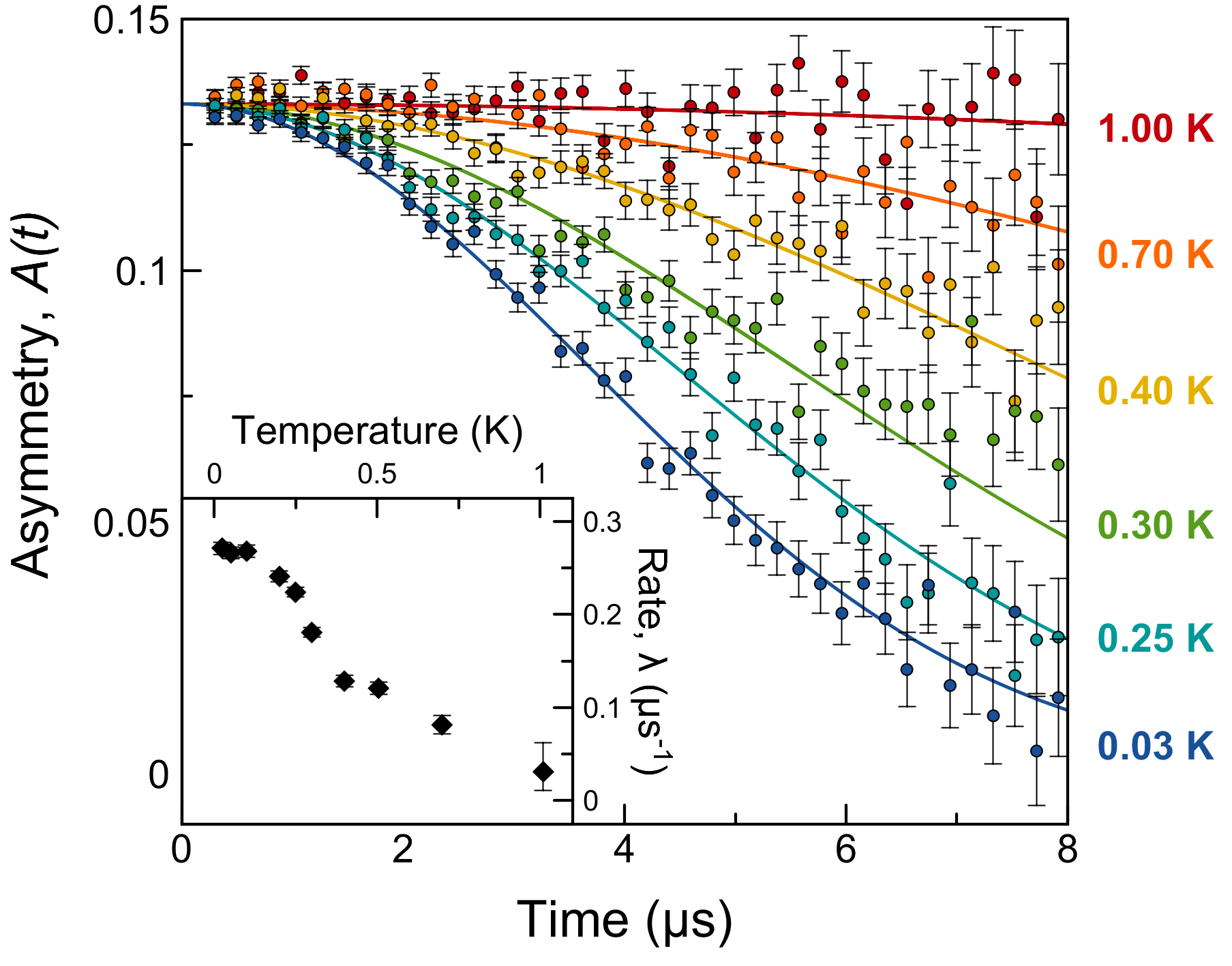}
\par
\caption{The $\mu$SR asymmetry spectra of Sm$_2$Ti$_2$O$_7$ between 1.00~K and 0.03~K, which are well-fit by a Gaussian relaxation, indicated by the solid lines. The temperature dependence of the relaxation rate, $\lambda$, extracted from these fits is shown in the inset. Below $T_N$, $\lambda$ is observed to plateau indicative of persistent spin dynamics in the ordered state. The absence of an oscillatory component in the asymmetry is consistent with moment fragmentation.}
\label{muSR}
\end{figure}

Last, we turn to zero-field muon spin relaxation ($\mu$SR) measurements on Sm$_2$Ti$_2$O$_7$, the results of which are presented in Fig.~\ref{muSR}. The temperature-independent contribution from muons that land outside the sample has been subtracted, leaving only the sample asymmetry. At 1~K and above, the asymmetry is non-relaxing, indicating that the Sm$^{3+}$ moments are in a fast-fluctuating paramagnetic regime. Approaching $T_N$, the relaxation gradually increases, consistent with a critical slowing of the spin dynamics. Over the full temperature range, the asymmetry is well-described by a Gaussian relaxation, $A(t)=A_0e^{-\lambda t^2}$, where $\lambda$ is the temperature-dependent relaxation rate. The fitted relaxation rates, which are weak at all temperatures, are plotted in the inset of Fig.~\ref{muSR} where we see the rate sharply increase at $T_N$ and then ultimately plateaus below 0.2~K. 

In a small moment sample such as Sm$_2$Ti$_2$O$_7$, where the background relaxation is weak, one would expect to observe spontaneous oscillations in the asymmetry spectra below $T_N$. However, they are strikingly absent in our measurement. The Gaussian relaxation observed here, combined with the lack of oscillations in the asymmetry spectra below $T_N$, is reminiscent of recent $\mu$SR measurements on another Ising antiferromagnet, Nd$_2$Zr$_2$O$_7$~\cite{xu2016spin}. In that case, the Gaussian relaxation was attributed to strong spin fluctuations that coexist microscopically with AIAO magnetic order, which generates a dynamic local magnetic field at the muon sites below $T_N$. This coexistance is argued to arise from magnetic moment fragmentation, which had been demonstrated in Nd$_2$Zr$_2$O$_7$ via neutron scattering~\cite{lhotel2015fluctuations,petit2016observation}. More specifically, the INS data on Nd$_2$Zr$_2$O$_7$ revealed that the dynamic component of the ground state has a characteristic frequency on the order of 10$^{10}$~Hz which is well within the $\mu$SR timescale. The persistent spin dynamics observed in our $\mu$SR spectra for Sm$_2$Ti$_2$O$_7$ could well arise from a similar origin. The absence of oscillations in an ordered state may also arise from a cancellation of the static dipolar field at the muon site from different ordered moments. However, this scenario is ruled out here by three simple observations: (1) there are no potential high-symmetry muon sites in the pyrochlore structure where the field could cancel by symmetry, (2) an oscillatory component \emph{has been} observed in the $\mu$SR of another AIAO pyrochlore Nd$_2$Sn$_2$O$_7$~\cite{bertin2015nd}, where it is important to note that, unlike Nd$_2$Zr$_2$O$_7$, fragmentation physics has not been demonstrated and (3) Sm$_2$Ti$_2$O$_7$ is iso-structural with Nd$_2$Sn$_2$O$_7$ and therefore the muon stopping sites are expected to be very similar.

We have demonstrated that Sm$_2$Ti$_2$O$_7$ possesses all the requisite ingredients for moment fragmentation physics. Crystal field analysis of our neutron spectroscopy measurements confirms that Sm$_2$Ti$_2$O$_7$ has an Ising dipolar-octupolar crystal field ground state doublet. Through symmetry analysis of our neutron diffraction data, we find that Sm$_2$Ti$_2$O$_7$ orders into an all-in, all-out magnetic structure below $T_N=0.35$~K, with an ordered moment of $\mu_{\text{ord}}=0.44(7)~\mu_B$. Muon spin relaxation measurements identify persistent spin dynamics to temperatures well-below $T_N$ and an absence of oscillations, consistent with a fragmentation scenario.

\begin{acknowledgments}

This research was supported by NSERC of Canada. A portion of this research used resources at the High Flux Isotope Reactor and Spallation Neutron Source, a DOE Office of Science User Facility operated by the Oak Ridge National Laboratory. CRW thanks the Canada Research Chair program (Tier II). CM thanks the Manitoba Government for support through the MGS. JAQ acknowledges technical support from M. Lacerte and S. Fortier and funding from FRQNT and CFREF. GS thanks J. Lin and A.T. Savici for useful discussions, and support in the analysis. HDZ acknowledges support from NSF DMR through Grant No. DMR-1350002. 

\end{acknowledgments}

\bibliography{Sm2Ti2O7_References}

\begin{thebibliography}{27}%
\makeatletter
\providecommand \@ifxundefined [1]{%
 \@ifx{#1\undefined}
}%
\providecommand \@ifnum [1]{%
 \ifnum #1\expandafter \@firstoftwo
 \else \expandafter \@secondoftwo
 \fi
}%
\providecommand \@ifx [1]{%
 \ifx #1\expandafter \@firstoftwo
 \else \expandafter \@secondoftwo
 \fi
}%
\providecommand \natexlab [1]{#1}%
\providecommand \enquote  [1]{``#1''}%
\providecommand \bibnamefont  [1]{#1}%
\providecommand \bibfnamefont [1]{#1}%
\providecommand \citenamefont [1]{#1}%
\providecommand \href@noop [0]{\@secondoftwo}%
\providecommand \href [0]{\begingroup \@sanitize@url \@href}%
\providecommand \@href[1]{\@@startlink{#1}\@@href}%
\providecommand \@@href[1]{\endgroup#1\@@endlink}%
\providecommand \@sanitize@url [0]{\catcode `\\12\catcode `\$12\catcode
  `\&12\catcode `\#12\catcode `\^12\catcode `\_12\catcode `\%12\relax}%
\providecommand \@@startlink[1]{}%
\providecommand \@@endlink[0]{}%
\providecommand \url  [0]{\begingroup\@sanitize@url \@url }%
\providecommand \@url [1]{\endgroup\@href {#1}{\urlprefix }}%
\providecommand \urlprefix  [0]{URL }%
\providecommand \Eprint [0]{\href }%
\providecommand \doibase [0]{http://dx.doi.org/}%
\providecommand \selectlanguage [0]{\@gobble}%
\providecommand \bibinfo  [0]{\@secondoftwo}%
\providecommand \bibfield  [0]{\@secondoftwo}%
\providecommand \translation [1]{[#1]}%
\providecommand \BibitemOpen [0]{}%
\providecommand \bibitemStop [0]{}%
\providecommand \bibitemNoStop [0]{.\EOS\space}%
\providecommand \EOS [0]{\spacefactor3000\relax}%
\providecommand \BibitemShut  [1]{\csname bibitem#1\endcsname}%
\let\auto@bib@innerbib\@empty
\bibitem [{\citenamefont {Gardner}\ \emph {et~al.}(2010)\citenamefont
  {Gardner}, \citenamefont {Gingras},\ and\ \citenamefont
  {Greedan}}]{gardner2010magnetic}%
  \BibitemOpen
  \bibfield  {author} {\bibinfo {author} {\bibfnamefont {Jason~S}\ \bibnamefont
  {Gardner}}, \bibinfo {author} {\bibfnamefont {Michel~JP}\ \bibnamefont
  {Gingras}}, \ and\ \bibinfo {author} {\bibfnamefont {John~E}\ \bibnamefont
  {Greedan}},\ }\bibfield  {title} {\enquote {\bibinfo {title} {Magnetic
  pyrochlore oxides},}\ }\href@noop {} {\bibfield  {journal} {\bibinfo
  {journal} {Reviews of Modern Physics}\ }\textbf {\bibinfo {volume} {82}},\
  \bibinfo {pages} {53} (\bibinfo {year} {2010})}\BibitemShut {NoStop}%
\bibitem [{\citenamefont {Harris}\ \emph {et~al.}(1997)\citenamefont {Harris},
  \citenamefont {Bramwell}, \citenamefont {McMorrow}, \citenamefont {Zeiske},\
  and\ \citenamefont {Godfrey}}]{harris1997geometrical}%
  \BibitemOpen
  \bibfield  {author} {\bibinfo {author} {\bibfnamefont {MJ}~\bibnamefont
  {Harris}}, \bibinfo {author} {\bibfnamefont {ST}~\bibnamefont {Bramwell}},
  \bibinfo {author} {\bibfnamefont {DF}~\bibnamefont {McMorrow}}, \bibinfo
  {author} {\bibfnamefont {Th}~\bibnamefont {Zeiske}}, \ and\ \bibinfo {author}
  {\bibfnamefont {KW}~\bibnamefont {Godfrey}},\ }\bibfield  {title} {\enquote
  {\bibinfo {title} {Geometrical frustration in the ferromagnetic pyrochlore
  {Ho$_2$Ti$_2$O$_7$}},}\ }\href@noop {} {\bibfield  {journal} {\bibinfo
  {journal} {Physical Review Letters}\ }\textbf {\bibinfo {volume} {79}},\
  \bibinfo {pages} {2554} (\bibinfo {year} {1997})}\BibitemShut {NoStop}%
\bibitem [{\citenamefont {Ramirez}\ \emph {et~al.}(1999)\citenamefont
  {Ramirez}, \citenamefont {Hayashi}, \citenamefont {Cava}, \citenamefont
  {Siddharthan},\ and\ \citenamefont {Shastry}}]{ramirez1999zero}%
  \BibitemOpen
  \bibfield  {author} {\bibinfo {author} {\bibfnamefont {Arthur~P}\
  \bibnamefont {Ramirez}}, \bibinfo {author} {\bibfnamefont {A}~\bibnamefont
  {Hayashi}}, \bibinfo {author} {\bibfnamefont {RJ}~\bibnamefont {Cava}},
  \bibinfo {author} {\bibfnamefont {R}~\bibnamefont {Siddharthan}}, \ and\
  \bibinfo {author} {\bibfnamefont {BS}~\bibnamefont {Shastry}},\ }\bibfield
  {title} {\enquote {\bibinfo {title} {Zero-point entropy in spin ice},}\
  }\href@noop {} {\bibfield  {journal} {\bibinfo  {journal} {Nature}\ }\textbf
  {\bibinfo {volume} {399}},\ \bibinfo {pages} {333} (\bibinfo {year}
  {1999})}\BibitemShut {NoStop}%
\bibitem [{\citenamefont {Bramwell}\ and\ \citenamefont
  {Gingras}(2001)}]{bramwell2001spin}%
  \BibitemOpen
  \bibfield  {author} {\bibinfo {author} {\bibfnamefont {Steven~T}\
  \bibnamefont {Bramwell}}\ and\ \bibinfo {author} {\bibfnamefont {Michel~JP}\
  \bibnamefont {Gingras}},\ }\bibfield  {title} {\enquote {\bibinfo {title}
  {Spin ice state in frustrated magnetic pyrochlore materials},}\ }\href@noop
  {} {\bibfield  {journal} {\bibinfo  {journal} {Science}\ }\textbf {\bibinfo
  {volume} {294}},\ \bibinfo {pages} {1495--1501} (\bibinfo {year}
  {2001})}\BibitemShut {NoStop}%
\bibitem [{\citenamefont {den Hertog}\ and\ \citenamefont
  {Gingras}(2000)}]{den2000dipolar}%
  \BibitemOpen
  \bibfield  {author} {\bibinfo {author} {\bibfnamefont {Byron~C}\ \bibnamefont
  {den Hertog}}\ and\ \bibinfo {author} {\bibfnamefont {Michel~JP}\
  \bibnamefont {Gingras}},\ }\bibfield  {title} {\enquote {\bibinfo {title}
  {Dipolar interactions and origin of spin ice in {Ising} pyrochlore
  magnets},}\ }\href@noop {} {\bibfield  {journal} {\bibinfo  {journal}
  {Physical Review Letters}\ }\textbf {\bibinfo {volume} {84}},\ \bibinfo
  {pages} {3430} (\bibinfo {year} {2000})}\BibitemShut {NoStop}%
\bibitem [{\citenamefont {Bertin}\ \emph {et~al.}(2015)\citenamefont {Bertin},
  \citenamefont {de~R{\'e}otier}, \citenamefont {F{\aa}k}, \citenamefont
  {Marin}, \citenamefont {Yaouanc}, \citenamefont {Forget}, \citenamefont
  {Sheptyakov}, \citenamefont {Frick}, \citenamefont {Ritter}, \citenamefont
  {Amato} \emph {et~al.}}]{bertin2015nd}%
  \BibitemOpen
  \bibfield  {author} {\bibinfo {author} {\bibfnamefont {Alexandre}\
  \bibnamefont {Bertin}}, \bibinfo {author} {\bibfnamefont {P~Dalmas}\
  \bibnamefont {de~R{\'e}otier}}, \bibinfo {author} {\bibfnamefont
  {B}~\bibnamefont {F{\aa}k}}, \bibinfo {author} {\bibfnamefont {Christophe}\
  \bibnamefont {Marin}}, \bibinfo {author} {\bibfnamefont {Alain}\ \bibnamefont
  {Yaouanc}}, \bibinfo {author} {\bibfnamefont {A}~\bibnamefont {Forget}},
  \bibinfo {author} {\bibfnamefont {D}~\bibnamefont {Sheptyakov}}, \bibinfo
  {author} {\bibfnamefont {Bernhard}\ \bibnamefont {Frick}}, \bibinfo {author}
  {\bibfnamefont {C}~\bibnamefont {Ritter}}, \bibinfo {author} {\bibfnamefont
  {A}~\bibnamefont {Amato}},  \emph {et~al.},\ }\bibfield  {title} {\enquote
  {\bibinfo {title} {{Nd$_2$Sn$_2$O$_7$}: {An} all-in--all-out pyrochlore
  magnet with no divergence-free field and anomalously slow paramagnetic spin
  dynamics},}\ }\href@noop {} {\bibfield  {journal} {\bibinfo  {journal}
  {Physical Review B}\ }\textbf {\bibinfo {volume} {92}},\ \bibinfo {pages}
  {144423} (\bibinfo {year} {2015})}\BibitemShut {NoStop}%
\bibitem [{\citenamefont {Xu}\ \emph {et~al.}(2015)\citenamefont {Xu},
  \citenamefont {Anand}, \citenamefont {Bera}, \citenamefont {Frontzek},
  \citenamefont {Abernathy}, \citenamefont {Casati}, \citenamefont
  {Siemensmeyer},\ and\ \citenamefont {Lake}}]{xu2015magnetic}%
  \BibitemOpen
  \bibfield  {author} {\bibinfo {author} {\bibfnamefont {J}~\bibnamefont {Xu}},
  \bibinfo {author} {\bibfnamefont {VK}~\bibnamefont {Anand}}, \bibinfo
  {author} {\bibfnamefont {AK}~\bibnamefont {Bera}}, \bibinfo {author}
  {\bibfnamefont {M}~\bibnamefont {Frontzek}}, \bibinfo {author} {\bibfnamefont
  {Douglas~L}\ \bibnamefont {Abernathy}}, \bibinfo {author} {\bibfnamefont
  {N}~\bibnamefont {Casati}}, \bibinfo {author} {\bibfnamefont {K}~\bibnamefont
  {Siemensmeyer}}, \ and\ \bibinfo {author} {\bibfnamefont {B}~\bibnamefont
  {Lake}},\ }\bibfield  {title} {\enquote {\bibinfo {title} {Magnetic structure
  and crystal-field states of the pyrochlore antiferromagnet
  {Nd$_2$Zr$_2$O$_7$}},}\ }\href@noop {} {\bibfield  {journal} {\bibinfo
  {journal} {Physical Review B}\ }\textbf {\bibinfo {volume} {92}},\ \bibinfo
  {pages} {224430} (\bibinfo {year} {2015})}\BibitemShut {NoStop}%
\bibitem [{\citenamefont {Lhotel}\ \emph {et~al.}(2015)\citenamefont {Lhotel},
  \citenamefont {Petit}, \citenamefont {Guitteny}, \citenamefont {Florea},
  \citenamefont {Hatnean}, \citenamefont {Colin}, \citenamefont {Ressouche},
  \citenamefont {Lees},\ and\ \citenamefont
  {Balakrishnan}}]{lhotel2015fluctuations}%
  \BibitemOpen
  \bibfield  {author} {\bibinfo {author} {\bibfnamefont {Elsa}\ \bibnamefont
  {Lhotel}}, \bibinfo {author} {\bibfnamefont {Sylvain}\ \bibnamefont {Petit}},
  \bibinfo {author} {\bibfnamefont {Sol{\`e}ne}\ \bibnamefont {Guitteny}},
  \bibinfo {author} {\bibfnamefont {O}~\bibnamefont {Florea}}, \bibinfo
  {author} {\bibfnamefont {M~Ciomaga}\ \bibnamefont {Hatnean}}, \bibinfo
  {author} {\bibfnamefont {Claire}\ \bibnamefont {Colin}}, \bibinfo {author}
  {\bibfnamefont {Eric}\ \bibnamefont {Ressouche}}, \bibinfo {author}
  {\bibfnamefont {MR}~\bibnamefont {Lees}}, \ and\ \bibinfo {author}
  {\bibfnamefont {G}~\bibnamefont {Balakrishnan}},\ }\bibfield  {title}
  {\enquote {\bibinfo {title} {Fluctuations and all-in--all-out ordering in
  dipole-octupole {Nd$_2$Zr$_2$O$_7$}},}\ }\href@noop {} {\bibfield  {journal}
  {\bibinfo  {journal} {Physical Review Letters}\ }\textbf {\bibinfo {volume}
  {115}},\ \bibinfo {pages} {197202} (\bibinfo {year} {2015})}\BibitemShut
  {NoStop}%
\bibitem [{\citenamefont {Anand}\ \emph {et~al.}(2015)\citenamefont {Anand},
  \citenamefont {Bera}, \citenamefont {Xu}, \citenamefont
  {Herrmannsd{\"o}rfer}, \citenamefont {Ritter},\ and\ \citenamefont
  {Lake}}]{anand2015observation}%
  \BibitemOpen
  \bibfield  {author} {\bibinfo {author} {\bibfnamefont {VK}~\bibnamefont
  {Anand}}, \bibinfo {author} {\bibfnamefont {AK}~\bibnamefont {Bera}},
  \bibinfo {author} {\bibfnamefont {J}~\bibnamefont {Xu}}, \bibinfo {author}
  {\bibfnamefont {T}~\bibnamefont {Herrmannsd{\"o}rfer}}, \bibinfo {author}
  {\bibfnamefont {C}~\bibnamefont {Ritter}}, \ and\ \bibinfo {author}
  {\bibfnamefont {B}~\bibnamefont {Lake}},\ }\bibfield  {title} {\enquote
  {\bibinfo {title} {Observation of long-range magnetic ordering in pyrohafnate
  {Nd$_2$Hf$_2$O$_7$}: a neutron diffraction study},}\ }\href@noop {}
  {\bibfield  {journal} {\bibinfo  {journal} {Physical Review B}\ }\textbf
  {\bibinfo {volume} {92}},\ \bibinfo {pages} {184418} (\bibinfo {year}
  {2015})}\BibitemShut {NoStop}%
\bibitem [{\citenamefont {Petit}\ \emph {et~al.}(2016)\citenamefont {Petit},
  \citenamefont {Lhotel}, \citenamefont {Canals}, \citenamefont {Hatnean},
  \citenamefont {Ollivier}, \citenamefont {Mutka}, \citenamefont {Ressouche},
  \citenamefont {Wildes}, \citenamefont {Lees},\ and\ \citenamefont
  {Balakrishnan}}]{petit2016observation}%
  \BibitemOpen
  \bibfield  {author} {\bibinfo {author} {\bibfnamefont {Sylvain}\ \bibnamefont
  {Petit}}, \bibinfo {author} {\bibfnamefont {Elsa}\ \bibnamefont {Lhotel}},
  \bibinfo {author} {\bibfnamefont {Benjamin}\ \bibnamefont {Canals}}, \bibinfo
  {author} {\bibfnamefont {M~Ciomaga}\ \bibnamefont {Hatnean}}, \bibinfo
  {author} {\bibfnamefont {Jacques}\ \bibnamefont {Ollivier}}, \bibinfo
  {author} {\bibfnamefont {Hannu}\ \bibnamefont {Mutka}}, \bibinfo {author}
  {\bibfnamefont {Eric}\ \bibnamefont {Ressouche}}, \bibinfo {author}
  {\bibfnamefont {AR}~\bibnamefont {Wildes}}, \bibinfo {author} {\bibfnamefont
  {MR}~\bibnamefont {Lees}}, \ and\ \bibinfo {author} {\bibfnamefont
  {G}~\bibnamefont {Balakrishnan}},\ }\bibfield  {title} {\enquote {\bibinfo
  {title} {Observation of magnetic fragmentation in spin ice},}\ }\href@noop {}
  {\bibfield  {journal} {\bibinfo  {journal} {Nature Physics}\ }\textbf
  {\bibinfo {volume} {12}},\ \bibinfo {pages} {746} (\bibinfo {year}
  {2016})}\BibitemShut {NoStop}%
\bibitem [{\citenamefont {Brooks-Bartlett}\ \emph {et~al.}(2014)\citenamefont
  {Brooks-Bartlett}, \citenamefont {Banks}, \citenamefont {Jaubert},
  \citenamefont {Harman-Clarke},\ and\ \citenamefont
  {Holdsworth}}]{brooks2014magnetic}%
  \BibitemOpen
  \bibfield  {author} {\bibinfo {author} {\bibfnamefont {ME}~\bibnamefont
  {Brooks-Bartlett}}, \bibinfo {author} {\bibfnamefont {Simon~T}\ \bibnamefont
  {Banks}}, \bibinfo {author} {\bibfnamefont {Ludovic~DC}\ \bibnamefont
  {Jaubert}}, \bibinfo {author} {\bibfnamefont {Adam}\ \bibnamefont
  {Harman-Clarke}}, \ and\ \bibinfo {author} {\bibfnamefont {Peter~CW}\
  \bibnamefont {Holdsworth}},\ }\bibfield  {title} {\enquote {\bibinfo {title}
  {Magnetic-moment fragmentation and monopole crystallization},}\ }\href@noop
  {} {\bibfield  {journal} {\bibinfo  {journal} {Physical Review X}\ }\textbf
  {\bibinfo {volume} {4}},\ \bibinfo {pages} {011007} (\bibinfo {year}
  {2014})}\BibitemShut {NoStop}%
\bibitem [{\citenamefont {Benton}(2016)}]{benton2016quantum}%
  \BibitemOpen
  \bibfield  {author} {\bibinfo {author} {\bibfnamefont {Owen}\ \bibnamefont
  {Benton}},\ }\bibfield  {title} {\enquote {\bibinfo {title} {Quantum origins
  of moment fragmentation in {Nd$_2$Zr$_2$O$_7$}},}\ }\href@noop {} {\bibfield
  {journal} {\bibinfo  {journal} {Physical Review B}\ }\textbf {\bibinfo
  {volume} {94}},\ \bibinfo {pages} {104430} (\bibinfo {year}
  {2016})}\BibitemShut {NoStop}%
\bibitem [{\citenamefont {Singh}\ \emph {et~al.}(2008)\citenamefont {Singh},
  \citenamefont {Saha}, \citenamefont {Dhar}, \citenamefont {Suryanarayanan},
  \citenamefont {Sood},\ and\ \citenamefont
  {Revcolevschi}}]{singh2008manifestation}%
  \BibitemOpen
  \bibfield  {author} {\bibinfo {author} {\bibfnamefont {Surjeet}\ \bibnamefont
  {Singh}}, \bibinfo {author} {\bibfnamefont {Surajit}\ \bibnamefont {Saha}},
  \bibinfo {author} {\bibfnamefont {SK}~\bibnamefont {Dhar}}, \bibinfo {author}
  {\bibfnamefont {R}~\bibnamefont {Suryanarayanan}}, \bibinfo {author}
  {\bibfnamefont {AK}~\bibnamefont {Sood}}, \ and\ \bibinfo {author}
  {\bibfnamefont {A}~\bibnamefont {Revcolevschi}},\ }\bibfield  {title}
  {\enquote {\bibinfo {title} {Manifestation of geometric frustration on
  magnetic and thermodynamic properties of the pyrochlores {Sm$_2$$X_2$O$_7$}
  ({$X$ = Ti, Zr})},}\ }\href@noop {} {\bibfield  {journal} {\bibinfo
  {journal} {Physical Review B}\ }\textbf {\bibinfo {volume} {77}},\ \bibinfo
  {pages} {054408} (\bibinfo {year} {2008})}\BibitemShut {NoStop}%
\bibitem [{\citenamefont {Abernathy}\ \emph {et~al.}(2012)\citenamefont
  {Abernathy}, \citenamefont {Stone}, \citenamefont {Loguillo}, \citenamefont
  {Lucas}, \citenamefont {Delaire}, \citenamefont {Tang}, \citenamefont {Lin},\
  and\ \citenamefont {Fultz}}]{abernathy2012design}%
  \BibitemOpen
  \bibfield  {author} {\bibinfo {author} {\bibfnamefont {Douglas~L}\
  \bibnamefont {Abernathy}}, \bibinfo {author} {\bibfnamefont {Matthew~B}\
  \bibnamefont {Stone}}, \bibinfo {author} {\bibfnamefont {MJ}~\bibnamefont
  {Loguillo}}, \bibinfo {author} {\bibfnamefont {MS}~\bibnamefont {Lucas}},
  \bibinfo {author} {\bibfnamefont {O}~\bibnamefont {Delaire}}, \bibinfo
  {author} {\bibfnamefont {Xiaoli}\ \bibnamefont {Tang}}, \bibinfo {author}
  {\bibfnamefont {JYY}\ \bibnamefont {Lin}}, \ and\ \bibinfo {author}
  {\bibfnamefont {B}~\bibnamefont {Fultz}},\ }\bibfield  {title} {\enquote
  {\bibinfo {title} {Design and operation of the wide angular-range chopper
  spectrometer {ARCS} at the {Spallation Neutron Source}},}\ }\href@noop {}
  {\bibfield  {journal} {\bibinfo  {journal} {Review of Scientific
  Instruments}\ }\textbf {\bibinfo {volume} {83}},\ \bibinfo {pages} {015114}
  (\bibinfo {year} {2012})}\BibitemShut {NoStop}%
\bibitem [{\citenamefont {Granroth}\ \emph {et~al.}(2010)\citenamefont
  {Granroth}, \citenamefont {Kolesnikov}, \citenamefont {Sherline},
  \citenamefont {Clancy}, \citenamefont {Ross}, \citenamefont {Ruff},
  \citenamefont {Gaulin},\ and\ \citenamefont {Nagler}}]{granroth2010sequoia}%
  \BibitemOpen
  \bibfield  {author} {\bibinfo {author} {\bibfnamefont {GE}~\bibnamefont
  {Granroth}}, \bibinfo {author} {\bibfnamefont {AI}~\bibnamefont
  {Kolesnikov}}, \bibinfo {author} {\bibfnamefont {TE}~\bibnamefont
  {Sherline}}, \bibinfo {author} {\bibfnamefont {JP}~\bibnamefont {Clancy}},
  \bibinfo {author} {\bibfnamefont {KA}~\bibnamefont {Ross}}, \bibinfo {author}
  {\bibfnamefont {JPC}\ \bibnamefont {Ruff}}, \bibinfo {author} {\bibfnamefont
  {BD}~\bibnamefont {Gaulin}}, \ and\ \bibinfo {author} {\bibfnamefont
  {SE}~\bibnamefont {Nagler}},\ }\bibfield  {title} {\enquote {\bibinfo {title}
  {{SEQUOIA}: a newly operating chopper spectrometer at the {SNS}},}\ }in\
  \href@noop {} {\emph {\bibinfo {booktitle} {Journal of Physics: Conference
  Series}}},\ Vol.\ \bibinfo {volume} {251}\ (\bibinfo {organization} {IOP
  Publishing},\ \bibinfo {year} {2010})\ p.\ \bibinfo {pages}
  {012058}\BibitemShut {NoStop}%
\bibitem [{\citenamefont {Malkin}\ \emph {et~al.}(2010)\citenamefont {Malkin},
  \citenamefont {Lummen}, \citenamefont {Van~Loosdrecht}, \citenamefont
  {Dhalenne},\ and\ \citenamefont {Zakirov}}]{malkin2010static}%
  \BibitemOpen
  \bibfield  {author} {\bibinfo {author} {\bibfnamefont {BZ}~\bibnamefont
  {Malkin}}, \bibinfo {author} {\bibfnamefont {TTA}\ \bibnamefont {Lummen}},
  \bibinfo {author} {\bibfnamefont {PHM}\ \bibnamefont {Van~Loosdrecht}},
  \bibinfo {author} {\bibfnamefont {G}~\bibnamefont {Dhalenne}}, \ and\
  \bibinfo {author} {\bibfnamefont {AR}~\bibnamefont {Zakirov}},\ }\bibfield
  {title} {\enquote {\bibinfo {title} {Static magnetic susceptibility, crystal
  field and exchange interactions in rare earth titanate pyrochlores},}\
  }\href@noop {} {\bibfield  {journal} {\bibinfo  {journal} {Journal of
  Physics: Condensed Matter}\ }\textbf {\bibinfo {volume} {22}},\ \bibinfo
  {pages} {276003} (\bibinfo {year} {2010})}\BibitemShut {NoStop}%
\bibitem [{\citenamefont {Lin}\ \emph {et~al.}(2016)\citenamefont {Lin},
  \citenamefont {Smith}, \citenamefont {Granroth}, \citenamefont {Abernathy},
  \citenamefont {Lumsden}, \citenamefont {Winn}, \citenamefont {Aczel},
  \citenamefont {Aivazis},\ and\ \citenamefont {Fultz}}]{lin2016mcvine}%
  \BibitemOpen
  \bibfield  {author} {\bibinfo {author} {\bibfnamefont {Jiao~YY}\ \bibnamefont
  {Lin}}, \bibinfo {author} {\bibfnamefont {Hillary~L}\ \bibnamefont {Smith}},
  \bibinfo {author} {\bibfnamefont {Garrett~E}\ \bibnamefont {Granroth}},
  \bibinfo {author} {\bibfnamefont {Douglas~L}\ \bibnamefont {Abernathy}},
  \bibinfo {author} {\bibfnamefont {Mark~D}\ \bibnamefont {Lumsden}}, \bibinfo
  {author} {\bibfnamefont {Barry}\ \bibnamefont {Winn}}, \bibinfo {author}
  {\bibfnamefont {Adam~A}\ \bibnamefont {Aczel}}, \bibinfo {author}
  {\bibfnamefont {Michael}\ \bibnamefont {Aivazis}}, \ and\ \bibinfo {author}
  {\bibfnamefont {Brent}\ \bibnamefont {Fultz}},\ }\bibfield  {title} {\enquote
  {\bibinfo {title} {{MCViNE}--an object oriented monte carlo neutron ray
  tracing simulation package},}\ }\href@noop {} {\bibfield  {journal} {\bibinfo
   {journal} {Nucl. Instr. Meth. Phys. Res.}\ }\textbf {\bibinfo {volume}
  {810}},\ \bibinfo {pages} {86--99} (\bibinfo {year} {2016})}\BibitemShut
  {NoStop}%
\bibitem [{\citenamefont {Blume}\ \emph {et~al.}(1964)\citenamefont {Blume},
  \citenamefont {Freeman},\ and\ \citenamefont {Watson}}]{blume1964theory}%
  \BibitemOpen
  \bibfield  {author} {\bibinfo {author} {\bibfnamefont {M}~\bibnamefont
  {Blume}}, \bibinfo {author} {\bibfnamefont {AJ}~\bibnamefont {Freeman}}, \
  and\ \bibinfo {author} {\bibfnamefont {RE}~\bibnamefont {Watson}},\
  }\bibfield  {title} {\enquote {\bibinfo {title} {Theory of spin-orbit
  coupling in atoms. {III}},}\ }\href@noop {} {\bibfield  {journal} {\bibinfo
  {journal} {Physical Review}\ }\textbf {\bibinfo {volume} {134}},\ \bibinfo
  {pages} {A320} (\bibinfo {year} {1964})}\BibitemShut {NoStop}%
\bibitem [{\citenamefont {Huang}\ \emph {et~al.}(2014)\citenamefont {Huang},
  \citenamefont {Chen},\ and\ \citenamefont {Hermele}}]{huang2014quantum}%
  \BibitemOpen
  \bibfield  {author} {\bibinfo {author} {\bibfnamefont {Yi-Ping}\ \bibnamefont
  {Huang}}, \bibinfo {author} {\bibfnamefont {Gang}\ \bibnamefont {Chen}}, \
  and\ \bibinfo {author} {\bibfnamefont {Michael}\ \bibnamefont {Hermele}},\
  }\bibfield  {title} {\enquote {\bibinfo {title} {Quantum spin ices and
  topological phases from dipolar-octupolar doublets on the pyrochlore
  lattice},}\ }\href@noop {} {\bibfield  {journal} {\bibinfo  {journal}
  {Physical Review Letters}\ }\textbf {\bibinfo {volume} {112}},\ \bibinfo
  {pages} {167203} (\bibinfo {year} {2014})}\BibitemShut {NoStop}%
\bibitem [{\citenamefont {Gaudet}\ \emph {et~al.}(2018)\citenamefont {Gaudet},
  \citenamefont {Hallas}, \citenamefont {Kolesnikov},\ and\ \citenamefont
  {Gaulin}}]{gaudet2017effect}%
  \BibitemOpen
  \bibfield  {author} {\bibinfo {author} {\bibfnamefont {J}~\bibnamefont
  {Gaudet}}, \bibinfo {author} {\bibfnamefont {AM}~\bibnamefont {Hallas}},
  \bibinfo {author} {\bibfnamefont {AI}~\bibnamefont {Kolesnikov}}, \ and\
  \bibinfo {author} {\bibfnamefont {BD}~\bibnamefont {Gaulin}},\ }\bibfield
  {title} {\enquote {\bibinfo {title} {Effect of chemical pressure on the
  crystal electric field states of erbium pyrochlore magnets},}\ }\href@noop {}
  {\bibfield  {journal} {\bibinfo  {journal} {Physical Review B}\ }\textbf
  {\bibinfo {volume} {97}},\ \bibinfo {pages} {024415} (\bibinfo {year}
  {2018})}\BibitemShut {NoStop}%
\bibitem [{\citenamefont {Gaudet}\ \emph {et~al.}(2015)\citenamefont {Gaudet},
  \citenamefont {Maharaj}, \citenamefont {Sala}, \citenamefont {Kermarrec},
  \citenamefont {Ross}, \citenamefont {Dabkowska}, \citenamefont {Kolesnikov},
  \citenamefont {Granroth},\ and\ \citenamefont {Gaulin}}]{gaudet2015neutron}%
  \BibitemOpen
  \bibfield  {author} {\bibinfo {author} {\bibfnamefont {J}~\bibnamefont
  {Gaudet}}, \bibinfo {author} {\bibfnamefont {DD}~\bibnamefont {Maharaj}},
  \bibinfo {author} {\bibfnamefont {G}~\bibnamefont {Sala}}, \bibinfo {author}
  {\bibfnamefont {E}~\bibnamefont {Kermarrec}}, \bibinfo {author}
  {\bibfnamefont {KA}~\bibnamefont {Ross}}, \bibinfo {author} {\bibfnamefont
  {HA}~\bibnamefont {Dabkowska}}, \bibinfo {author} {\bibfnamefont
  {AI}~\bibnamefont {Kolesnikov}}, \bibinfo {author} {\bibfnamefont
  {GE}~\bibnamefont {Granroth}}, \ and\ \bibinfo {author} {\bibfnamefont
  {BD}~\bibnamefont {Gaulin}},\ }\bibfield  {title} {\enquote {\bibinfo {title}
  {Neutron spectroscopic study of crystalline electric field excitations in
  stoichiometric and lightly stuffed {Yb$_2$Ti$_2$O$_7$}},}\ }\href@noop {}
  {\bibfield  {journal} {\bibinfo  {journal} {Physical Review B}\ }\textbf
  {\bibinfo {volume} {92}},\ \bibinfo {pages} {134420} (\bibinfo {year}
  {2015})}\BibitemShut {NoStop}%
\bibitem [{\citenamefont {Hutchings}(1964)}]{Hutchings1964227}%
  \BibitemOpen
  \bibfield  {author} {\bibinfo {author} {\bibfnamefont {M.T.}\ \bibnamefont
  {Hutchings}},\ }\bibfield  {title} {\enquote {\bibinfo {title} {Point-charge
  calculations of energy levels of magnetic ions in crystalline electric
  fields},}\ \ }(\bibinfo  {publisher} {Academic Press},\ \bibinfo {year}
  {1964})\ pp.\ \bibinfo {pages} {227 -- 273}\BibitemShut {NoStop}%
\bibitem [{\citenamefont {Rosenkranz}\ \emph {et~al.}(2000)\citenamefont
  {Rosenkranz}, \citenamefont {Ramirez}, \citenamefont {Hayashi}, \citenamefont
  {Cava}, \citenamefont {Siddharthan},\ and\ \citenamefont
  {Shastry}}]{rosenkranz2000crystal}%
  \BibitemOpen
  \bibfield  {author} {\bibinfo {author} {\bibfnamefont {S}~\bibnamefont
  {Rosenkranz}}, \bibinfo {author} {\bibfnamefont {AP}~\bibnamefont {Ramirez}},
  \bibinfo {author} {\bibfnamefont {A}~\bibnamefont {Hayashi}}, \bibinfo
  {author} {\bibfnamefont {RJ}~\bibnamefont {Cava}}, \bibinfo {author}
  {\bibfnamefont {R}~\bibnamefont {Siddharthan}}, \ and\ \bibinfo {author}
  {\bibfnamefont {BS}~\bibnamefont {Shastry}},\ }\bibfield  {title} {\enquote
  {\bibinfo {title} {Crystal-field interaction in the pyrochlore magnet
  {Ho$_2$Ti$_2$O$_7$}},}\ }\href@noop {} {\bibfield  {journal} {\bibinfo
  {journal} {Journal of Applied Physics}\ }\textbf {\bibinfo {volume} {87}},\
  \bibinfo {pages} {5914--5916} (\bibinfo {year} {2000})}\BibitemShut {NoStop}%
\bibitem [{\citenamefont {Bertin}\ \emph {et~al.}(2012)\citenamefont {Bertin},
  \citenamefont {Chapuis}, \citenamefont {de~R{\'e}otier},\ and\ \citenamefont
  {Yaouanc}}]{bertin2012crystal}%
  \BibitemOpen
  \bibfield  {author} {\bibinfo {author} {\bibfnamefont {A}~\bibnamefont
  {Bertin}}, \bibinfo {author} {\bibfnamefont {Y}~\bibnamefont {Chapuis}},
  \bibinfo {author} {\bibfnamefont {P~Dalmas}\ \bibnamefont {de~R{\'e}otier}},
  \ and\ \bibinfo {author} {\bibfnamefont {A}~\bibnamefont {Yaouanc}},\
  }\bibfield  {title} {\enquote {\bibinfo {title} {Crystal electric field in
  the {$R_2$Ti$_2$O$_7$} pyrochlore compounds},}\ }\href@noop {} {\bibfield
  {journal} {\bibinfo  {journal} {Journal of Physics: Condensed Matter}\
  }\textbf {\bibinfo {volume} {24}},\ \bibinfo {pages} {256003} (\bibinfo
  {year} {2012})}\BibitemShut {NoStop}%
\bibitem [{\citenamefont {Ruminy}\ \emph {et~al.}(2016)\citenamefont {Ruminy},
  \citenamefont {Pomjakushina}, \citenamefont {Iida}, \citenamefont {Kamazawa},
  \citenamefont {Adroja}, \citenamefont {Stuhr},\ and\ \citenamefont
  {Fennell}}]{ruminy2016crystal}%
  \BibitemOpen
  \bibfield  {author} {\bibinfo {author} {\bibfnamefont {M}~\bibnamefont
  {Ruminy}}, \bibinfo {author} {\bibfnamefont {E}~\bibnamefont {Pomjakushina}},
  \bibinfo {author} {\bibfnamefont {K}~\bibnamefont {Iida}}, \bibinfo {author}
  {\bibfnamefont {K}~\bibnamefont {Kamazawa}}, \bibinfo {author} {\bibfnamefont
  {DT}~\bibnamefont {Adroja}}, \bibinfo {author} {\bibfnamefont
  {U}~\bibnamefont {Stuhr}}, \ and\ \bibinfo {author} {\bibfnamefont
  {T}~\bibnamefont {Fennell}},\ }\bibfield  {title} {\enquote {\bibinfo {title}
  {Crystal-field parameters of the rare-earth pyrochlores {$R_2$Ti$_2$O$_7$}
  {($R$ = Tb, Dy, and Ho)}},}\ }\href@noop {} {\bibfield  {journal} {\bibinfo
  {journal} {Physical Review B}\ }\textbf {\bibinfo {volume} {94}},\ \bibinfo
  {pages} {024430} (\bibinfo {year} {2016})}\BibitemShut {NoStop}%
\bibitem [{\citenamefont {Princep}\ \emph {et~al.}(2015)\citenamefont
  {Princep}, \citenamefont {Walker}, \citenamefont {Adroja}, \citenamefont
  {Prabhakaran},\ and\ \citenamefont {Boothroyd}}]{princep2015crystal}%
  \BibitemOpen
  \bibfield  {author} {\bibinfo {author} {\bibfnamefont {AJ}~\bibnamefont
  {Princep}}, \bibinfo {author} {\bibfnamefont {HC}~\bibnamefont {Walker}},
  \bibinfo {author} {\bibfnamefont {DT}~\bibnamefont {Adroja}}, \bibinfo
  {author} {\bibfnamefont {D}~\bibnamefont {Prabhakaran}}, \ and\ \bibinfo
  {author} {\bibfnamefont {AT}~\bibnamefont {Boothroyd}},\ }\bibfield  {title}
  {\enquote {\bibinfo {title} {Crystal field states of {Tb$^{3+}$} in the
  pyrochlore spin liquid {Tb$_2$Ti$_2$O$_7$} from neutron spectroscopy},}\
  }\href@noop {} {\bibfield  {journal} {\bibinfo  {journal} {Physical Review
  B}\ }\textbf {\bibinfo {volume} {91}},\ \bibinfo {pages} {224430} (\bibinfo
  {year} {2015})}\BibitemShut {NoStop}%
\bibitem [{\citenamefont {Xu}\ \emph {et~al.}(2016)\citenamefont {Xu},
  \citenamefont {Balz}, \citenamefont {Baines}, \citenamefont {Luetkens},\ and\
  \citenamefont {Lake}}]{xu2016spin}%
  \BibitemOpen
  \bibfield  {author} {\bibinfo {author} {\bibfnamefont {J}~\bibnamefont {Xu}},
  \bibinfo {author} {\bibfnamefont {C}~\bibnamefont {Balz}}, \bibinfo {author}
  {\bibfnamefont {C}~\bibnamefont {Baines}}, \bibinfo {author} {\bibfnamefont
  {H}~\bibnamefont {Luetkens}}, \ and\ \bibinfo {author} {\bibfnamefont
  {B}~\bibnamefont {Lake}},\ }\bibfield  {title} {\enquote {\bibinfo {title}
  {Spin dynamics of the ordered dipolar-octupolar pseudospin-{1/2} pyrochlore
  {Nd$_2$Zr$_2$O$_7$} probed by muon spin relaxation},}\ }\href@noop {}
  {\bibfield  {journal} {\bibinfo  {journal} {Physical Review B}\ }\textbf
  {\bibinfo {volume} {94}},\ \bibinfo {pages} {064425} (\bibinfo {year}
  {2016})}\BibitemShut {NoStop}%
\end{thebibliography}%


\begin{thebibliography}{10}%
\makeatletter
\providecommand \@ifxundefined [1]{%
 \@ifx{#1\undefined}
}%
\providecommand \@ifnum [1]{%
 \ifnum #1\expandafter \@firstoftwo
 \else \expandafter \@secondoftwo
 \fi
}%
\providecommand \@ifx [1]{%
 \ifx #1\expandafter \@firstoftwo
 \else \expandafter \@secondoftwo
 \fi
}%
\providecommand \natexlab [1]{#1}%
\providecommand \enquote  [1]{``#1''}%
\providecommand \bibnamefont  [1]{#1}%
\providecommand \bibfnamefont [1]{#1}%
\providecommand \citenamefont [1]{#1}%
\providecommand \href@noop [0]{\@secondoftwo}%
\providecommand \href [0]{\begingroup \@sanitize@url \@href}%
\providecommand \@href[1]{\@@startlink{#1}\@@href}%
\providecommand \@@href[1]{\endgroup#1\@@endlink}%
\providecommand \@sanitize@url [0]{\catcode `\\12\catcode `\$12\catcode
  `\&12\catcode `\#12\catcode `\^12\catcode `\_12\catcode `\%12\relax}%
\providecommand \@@startlink[1]{}%
\providecommand \@@endlink[0]{}%
\providecommand \url  [0]{\begingroup\@sanitize@url \@url }%
\providecommand \@url [1]{\endgroup\@href {#1}{\urlprefix }}%
\providecommand \urlprefix  [0]{URL }%
\providecommand \Eprint [0]{\href }%
\providecommand \doibase [0]{http://dx.doi.org/}%
\providecommand \selectlanguage [0]{\@gobble}%
\providecommand \bibinfo  [0]{\@secondoftwo}%
\providecommand \bibfield  [0]{\@secondoftwo}%
\providecommand \translation [1]{[#1]}%
\providecommand \BibitemOpen [0]{}%
\providecommand \bibitemStop [0]{}%
\providecommand \bibitemNoStop [0]{.\EOS\space}%
\providecommand \EOS [0]{\spacefactor3000\relax}%
\providecommand \BibitemShut  [1]{\csname bibitem#1\endcsname}%
\let\auto@bib@innerbib\@empty
\bibitem [{\citenamefont {Lin}\ \emph {et~al.}(2016)\citenamefont {Lin},
  \citenamefont {Smith}, \citenamefont {Granroth}, \citenamefont {Abernathy},
  \citenamefont {Lumsden}, \citenamefont {Winn}, \citenamefont {Aczel},
  \citenamefont {Aivazis},\ and\ \citenamefont {Fultz}}]{lin2016mcvine}%
  \BibitemOpen
  \bibfield  {author} {\bibinfo {author} {\bibfnamefont {Jiao~YY}\ \bibnamefont
  {Lin}}, \bibinfo {author} {\bibfnamefont {Hillary~L}\ \bibnamefont {Smith}},
  \bibinfo {author} {\bibfnamefont {Garrett~E}\ \bibnamefont {Granroth}},
  \bibinfo {author} {\bibfnamefont {Douglas~L}\ \bibnamefont {Abernathy}},
  \bibinfo {author} {\bibfnamefont {Mark~D}\ \bibnamefont {Lumsden}}, \bibinfo
  {author} {\bibfnamefont {Barry}\ \bibnamefont {Winn}}, \bibinfo {author}
  {\bibfnamefont {Adam~A}\ \bibnamefont {Aczel}}, \bibinfo {author}
  {\bibfnamefont {Michael}\ \bibnamefont {Aivazis}}, \ and\ \bibinfo {author}
  {\bibfnamefont {Brent}\ \bibnamefont {Fultz}},\ }\bibfield  {title} {\enquote
  {\bibinfo {title} {{MCViNE}--an object oriented monte carlo neutron ray
  tracing simulation package},}\ }\href@noop {} {\bibfield  {journal} {\bibinfo
   {journal} {Nucl. Instr. Meth. Phys. Res.}\ }\textbf {\bibinfo {volume}
  {810}},\ \bibinfo {pages} {86--99} (\bibinfo {year} {2016})}\BibitemShut
  {NoStop}%
\bibitem [{\citenamefont {Wills}(2000)}]{wills2000new}%
  \BibitemOpen
  \bibfield  {author} {\bibinfo {author} {\bibfnamefont {AS}~\bibnamefont
  {Wills}},\ }\bibfield  {title} {\enquote {\bibinfo {title} {A new protocol
  for the determination of magnetic structures using simulated annealing and
  representational analysis ({SARAh})},}\ }\href@noop {} {\bibfield  {journal}
  {\bibinfo  {journal} {Physica B: Condensed Matter}\ }\textbf {\bibinfo
  {volume} {276}},\ \bibinfo {pages} {680--681} (\bibinfo {year}
  {2000})}\BibitemShut {NoStop}%
\bibitem [{\citenamefont
  {Rodr{\'\i}guez-Carvajal}(1993)}]{rodriguez1993recent}%
  \BibitemOpen
  \bibfield  {author} {\bibinfo {author} {\bibfnamefont {Juan}\ \bibnamefont
  {Rodr{\'\i}guez-Carvajal}},\ }\bibfield  {title} {\enquote {\bibinfo {title}
  {Recent advances in magnetic structure determination by neutron powder
  diffraction},}\ }\href@noop {} {\bibfield  {journal} {\bibinfo  {journal}
  {Physica B: Condensed Matter}\ }\textbf {\bibinfo {volume} {192}},\ \bibinfo
  {pages} {55--69} (\bibinfo {year} {1993})}\BibitemShut {NoStop}%
\bibitem [{\citenamefont {Ruminy}\ \emph {et~al.}(2016)\citenamefont {Ruminy},
  \citenamefont {Valdez}, \citenamefont {Wehinger}, \citenamefont {Bosak},
  \citenamefont {Adroja}, \citenamefont {Stuhr}, \citenamefont {Iida},
  \citenamefont {Kamazawa}, \citenamefont {Pomjakushina}, \citenamefont
  {Prabakharan} \emph {et~al.}}]{ruminy2016first}%
  \BibitemOpen
  \bibfield  {author} {\bibinfo {author} {\bibfnamefont {M}~\bibnamefont
  {Ruminy}}, \bibinfo {author} {\bibfnamefont {M~N{\'u}{\~n}ez}\ \bibnamefont
  {Valdez}}, \bibinfo {author} {\bibfnamefont {Bj{\"o}rn}\ \bibnamefont
  {Wehinger}}, \bibinfo {author} {\bibfnamefont {A}~\bibnamefont {Bosak}},
  \bibinfo {author} {\bibfnamefont {DT}~\bibnamefont {Adroja}}, \bibinfo
  {author} {\bibfnamefont {U}~\bibnamefont {Stuhr}}, \bibinfo {author}
  {\bibfnamefont {K}~\bibnamefont {Iida}}, \bibinfo {author} {\bibfnamefont
  {K}~\bibnamefont {Kamazawa}}, \bibinfo {author} {\bibfnamefont
  {E}~\bibnamefont {Pomjakushina}}, \bibinfo {author} {\bibfnamefont
  {D}~\bibnamefont {Prabakharan}},  \emph {et~al.},\ }\bibfield  {title}
  {\enquote {\bibinfo {title} {First-principles calculation and experimental
  investigation of lattice dynamics in the rare-earth pyrochlores
  {$R_2$Ti$_2$O$_7$ ($R$ = Tb, Dy, Ho)}},}\ }\href@noop {} {\bibfield
  {journal} {\bibinfo  {journal} {Physical Review B}\ }\textbf {\bibinfo
  {volume} {93}},\ \bibinfo {pages} {214308} (\bibinfo {year}
  {2016})}\BibitemShut {NoStop}%
\bibitem [{\citenamefont {Hutchings}(1964)}]{Hutchings1964227}%
  \BibitemOpen
  \bibfield  {author} {\bibinfo {author} {\bibfnamefont {M.T.}\ \bibnamefont
  {Hutchings}},\ }\bibfield  {title} {\enquote {\bibinfo {title} {Point-charge
  calculations of energy levels of magnetic ions in crystalline electric
  fields},}\ \ }(\bibinfo  {publisher} {Academic Press},\ \bibinfo {year}
  {1964})\ pp.\ \bibinfo {pages} {227 -- 273}\BibitemShut {NoStop}%
\bibitem [{\citenamefont {Stevens}(1952)}]{stevens1952matrix}%
  \BibitemOpen
  \bibfield  {author} {\bibinfo {author} {\bibfnamefont {KWH}\ \bibnamefont
  {Stevens}},\ }\bibfield  {title} {\enquote {\bibinfo {title} {Matrix elements
  and operator equivalents connected with the magnetic properties of rare earth
  ions},}\ }\href@noop {} {\bibfield  {journal} {\bibinfo  {journal}
  {Proceedings of the Physical Society. Section A}\ }\textbf {\bibinfo {volume}
  {65}},\ \bibinfo {pages} {209} (\bibinfo {year} {1952})}\BibitemShut
  {NoStop}%
\bibitem [{\citenamefont {Walter}(1984)}]{walter1984treating}%
  \BibitemOpen
  \bibfield  {author} {\bibinfo {author} {\bibfnamefont {U}~\bibnamefont
  {Walter}},\ }\bibfield  {title} {\enquote {\bibinfo {title} {Treating crystal
  field parameters in lower than cubic symmetries},}\ }\href@noop {} {\bibfield
   {journal} {\bibinfo  {journal} {Journal of Physics and Chemistry of Solids}\
  }\textbf {\bibinfo {volume} {45}},\ \bibinfo {pages} {401--408} (\bibinfo
  {year} {1984})}\BibitemShut {NoStop}%
\bibitem [{\citenamefont {Prather}(1961)}]{prather1961atomic}%
  \BibitemOpen
  \bibfield  {author} {\bibinfo {author} {\bibfnamefont {John~L}\ \bibnamefont
  {Prather}},\ }\href@noop {} {\emph {\bibinfo {title} {Atomic energy levels in
  crystals}}},\ \bibinfo {type} {Tech. Rep.}\ (\bibinfo  {institution}
  {National Bureau of Standards, Gaithersburg MD},\ \bibinfo {year}
  {1961})\BibitemShut {NoStop}%
\bibitem [{\citenamefont {Gaudet}\ \emph {et~al.}(2018)\citenamefont {Gaudet},
  \citenamefont {Hallas}, \citenamefont {Kolesnikov},\ and\ \citenamefont
  {Gaulin}}]{gaudet2017effect}%
  \BibitemOpen
  \bibfield  {author} {\bibinfo {author} {\bibfnamefont {J}~\bibnamefont
  {Gaudet}}, \bibinfo {author} {\bibfnamefont {AM}~\bibnamefont {Hallas}},
  \bibinfo {author} {\bibfnamefont {AI}~\bibnamefont {Kolesnikov}}, \ and\
  \bibinfo {author} {\bibfnamefont {BD}~\bibnamefont {Gaulin}},\ }\bibfield
  {title} {\enquote {\bibinfo {title} {Effect of chemical pressure on the
  crystal electric field states of erbium pyrochlore magnets},}\ }\href@noop {}
  {\bibfield  {journal} {\bibinfo  {journal} {Physical Review B}\ }\textbf
  {\bibinfo {volume} {97}},\ \bibinfo {pages} {024415} (\bibinfo {year}
  {2018})}\BibitemShut {NoStop}%
\bibitem [{\citenamefont {Knop}\ \emph {et~al.}(1969)\citenamefont {Knop},
  \citenamefont {Brisse},\ and\ \citenamefont
  {Castelliz}}]{knop1969pyrochlores}%
  \BibitemOpen
  \bibfield  {author} {\bibinfo {author} {\bibfnamefont {Osvald}\ \bibnamefont
  {Knop}}, \bibinfo {author} {\bibfnamefont {Fran{\c{c}}ois}\ \bibnamefont
  {Brisse}}, \ and\ \bibinfo {author} {\bibfnamefont {Lotte}\ \bibnamefont
  {Castelliz}},\ }\bibfield  {title} {\enquote {\bibinfo {title} {Pyrochlores
  {V}: Thermoanalytic, x-ray, neutron, infrared, and dielectric studies of
  {$A_2$Ti$_2$O$_7$} titanates},}\ }\href@noop {} {\bibfield  {journal}
  {\bibinfo  {journal} {Canadian Journal of Chemistry}\ }\textbf {\bibinfo
  {volume} {47}},\ \bibinfo {pages} {971--990} (\bibinfo {year}
  {1969})}\BibitemShut {NoStop}%
\end{thebibliography}%


\end{document}


\title{SUPPLEMENTAL MATERIAL \\
Dipolar-Octupolar Ising Antiferromagnetism in Sm$_2$Ti$_2$O$_7$:\\ A Moment Fragmentation Candidate}

\author{C. Mauws}
\affiliation{Department of Chemistry, University of Manitoba, Winnipeg R3T 2N2, Canada}
\affiliation{Department of Chemistry, University of Winnipeg, Winnipeg R3B 2E9, Canada}

\author{A.~M.~Hallas}
\affiliation{Department of Physics and Astronomy, McMaster University, Hamilton L8S 4M1, Canada}
\affiliation{Department of Physics and Astronomy and Rice Center for Quantum Materials, Rice University, Houston, TX, 77005 USA}

\author{G.~Sala}
\affiliation{Neutron Scattering Division, Oak Ridge National Laboratory, Oak Ridge, Tennessee 37831, USA}

\author{A.~A.~Aczel}
\affiliation{Neutron Scattering Division, Oak Ridge National Laboratory, Oak Ridge, Tennessee 37831, USA}

\author{P.~M.~Sarte}
\affiliation{School of Chemistry, University of Edinburgh, Edinburgh EH9 3FJ, United Kingdom}
\affiliation{Centre for Science at Extreme Conditions, University of Edinburgh, Edinburgh EH9 3FD, United Kingdom}

\author{J.~Gaudet}
\affiliation{Department of Physics and Astronomy, McMaster University, Hamilton L8S 4M1, Canada}

\author{D.~Ziat}
\affiliation{Institut Quantique and D\'{e}partement de Physique, Universit\'{e} de Sherbrooke, Sherbrooke, Qu\'{e}bec J1K 2R1, Canada}

\author{J.~A.~Quilliam}
\affiliation{Institut Quantique and D\'{e}partement de Physique, Universit\'{e} de Sherbrooke, Sherbrooke, Qu\'{e}bec J1K 2R1, Canada}

\author{J.~A.~Lussier}
\affiliation{Department of Chemistry, University of Manitoba, Winnipeg R3T 2N2, Canada}

\author{M. Bieringer}
\affiliation{Department of Chemistry, University of Manitoba, Winnipeg R3T 2N2, Canada}

\author{H.~D. Zhou}
\affiliation{Department of Physics and Astronomy, University of Tennessee-Knoxville, Knoxville 37996-1220, United States}
\affiliation{National High Magnetic Field Laboratory, Florida State University, Tallahassee 32306-4005, United States}

\author{A.~Wildes}
\affiliation {Institut Laue-Langevin, 71 avenue des Martyrs, CS 20156, 38042 Grenoble Cedex 9, France}

\author{M.~B.~Stone}
\affiliation{Neutron Scattering Division, Oak Ridge National Laboratory, Oak Ridge, Tennessee 37831, USA}

\author{D.~Abernathy}
\affiliation{Neutron Scattering Division, Oak Ridge National Laboratory, Oak Ridge, Tennessee 37831, USA}

\author{G.~M.~Luke}
\affiliation{Department of Physics and Astronomy, McMaster University, Hamilton L8S 4M1, Canada}
\affiliation{Canadian Institute for Advanced Research, Toronto M5G 1M1, Canada}
\affiliation{TRIUMF, 4004 Wesbrook Mall, Vancouver, British Columbia, Canada V6T 2A3}

\author{B.~D.~Gaulin}
\affiliation{Department of Physics and Astronomy, McMaster University, Hamilton L8S 4M1, Canada}
\affiliation{Canadian Institute for Advanced Research, Toronto M5G 1M1, Canada}

\author{C.~R.~Wiebe}
\affiliation{Department of Chemistry, University of Manitoba, Winnipeg R3T 2N2, Canada}
\affiliation{Department of Chemistry, University of Winnipeg, Winnipeg R3B 2E9, Canada}
\affiliation{Department of Physics and Astronomy, McMaster University, Hamilton L8S 4M1, Canada}
\affiliation{Canadian Institute for Advanced Research, Toronto M5G 1M1, Canada}

\date{\today}

\begin{abstract}
In this supplementary information we present the experimental details for the neutron scattering and muon spin resonance experiments on Sm$_2$Ti$_2$O$_7$. We then describe the Monte Carlo simulation (MC) we performed using MCViNE~\cite{lin2016mcvine} in order to estimate the absorption correction to the neutron spectroscopy of Sm$^{3+}$ in Sm$_2$Ti$_2$O$_7$.  Finally, we discuss the details of the crystal field calculation which we used to model the inelastic neutron scattering data collected at the ARCS spectrometer.  
\end{abstract}

\maketitle


\subsection{\label{sec: 0} Experimental Details}

Polycrystalline samples of Sm$_2$Ti$_2$O$_7$ were prepared by conventional solid state synthesis, with repeated firings at 1450 C until phase pure. Multiple single crystals were grown in a Quantum Design image furnace to optimize growth conditions. Growth under flowing argon yielded the best samples, judged by the homogeneous light yellow hue, visually similar to the powder, as opposed to other darker samples. Due to the high absorption cross section of natural abundance samarium, a single crystal was grown using 99.8\% enriched $^{154}$Sm$_2$O$_3$ (Cambridge Isotopes), under flowing argon gas. 

Low-temperature specific heat measurements were performed using the quasi-adiabatic technique. The heater and thermometer were directly attached to a single crystal plaquette of Sm$_2$Ti$_2$O$_7$ which was weakly linked to the mixing chamber of a dilution refrigerator. 

An approximately 3 gram segment of the isotopically enriched Sm$_2$Ti$_2$O$_7$ crystal was used for neutron diffraction measurements on D7 at the Institut Laue-Langevin (ILL) and on HB-1A at the High Flux Isotope Reactor at the Oak Ridge National Laboratory (ORNL). The crystal was aligned in the (HHL) plane, attached to a copper mount with copper wire to ensure good thermal contact, and then loaded in a dilution fridge insert with a base temperature of 50 mK for these experiments. A second segment of the crystal was used for crystal field (CEF) measurements on SEQUOIA at the Spallation Neutron Source at ORNL. The crystal was mounted on an aluminum plate with the (HHL) plane horizontal and then loaded in a cryostat with a base temperature of 1.8 K. The crystal field measurements were later reproduced on the ARCS spectrometer, to provide improved high-temperature data. As performing an empty can subtraction on SEQUOIA was difficult due to the absorption of the sample, for the ARCS experiment the crystal was simply attached to a minimal amount of aluminum wire such that no empty can subtraction was required.  Symmetry analysis was performed with SARAh~\cite{wills2000new} and the relative peak intensities were calculated with the FullProf Suite \cite{rodriguez1993recent}. 

Zero-field muon spin relaxation (ZF-$\mu$SR) measurements were performed with the Pandora spectrometer at the M15 muon beam line at TRIUMF. A single crystal of non-isotopically enriched Sm$_2$Ti$_2$O$_7$ was aligned along the [001] direction and cut into slices approximately 2 mm thick covering a total surface area of 2 cm$^2$. These crystal slices were affixed directly to the dilution refrigerator cold finger using Apiezon n-grease. Muons, initially polarized anti-parallel to their momentum, were incident upon the [001] face. 

\subsection{\label{sec: 1} Monte Carlo Simulation using MCViNE} 
        
\begin{figure}[tbp]
\centering
\includegraphics[width=1\columnwidth]{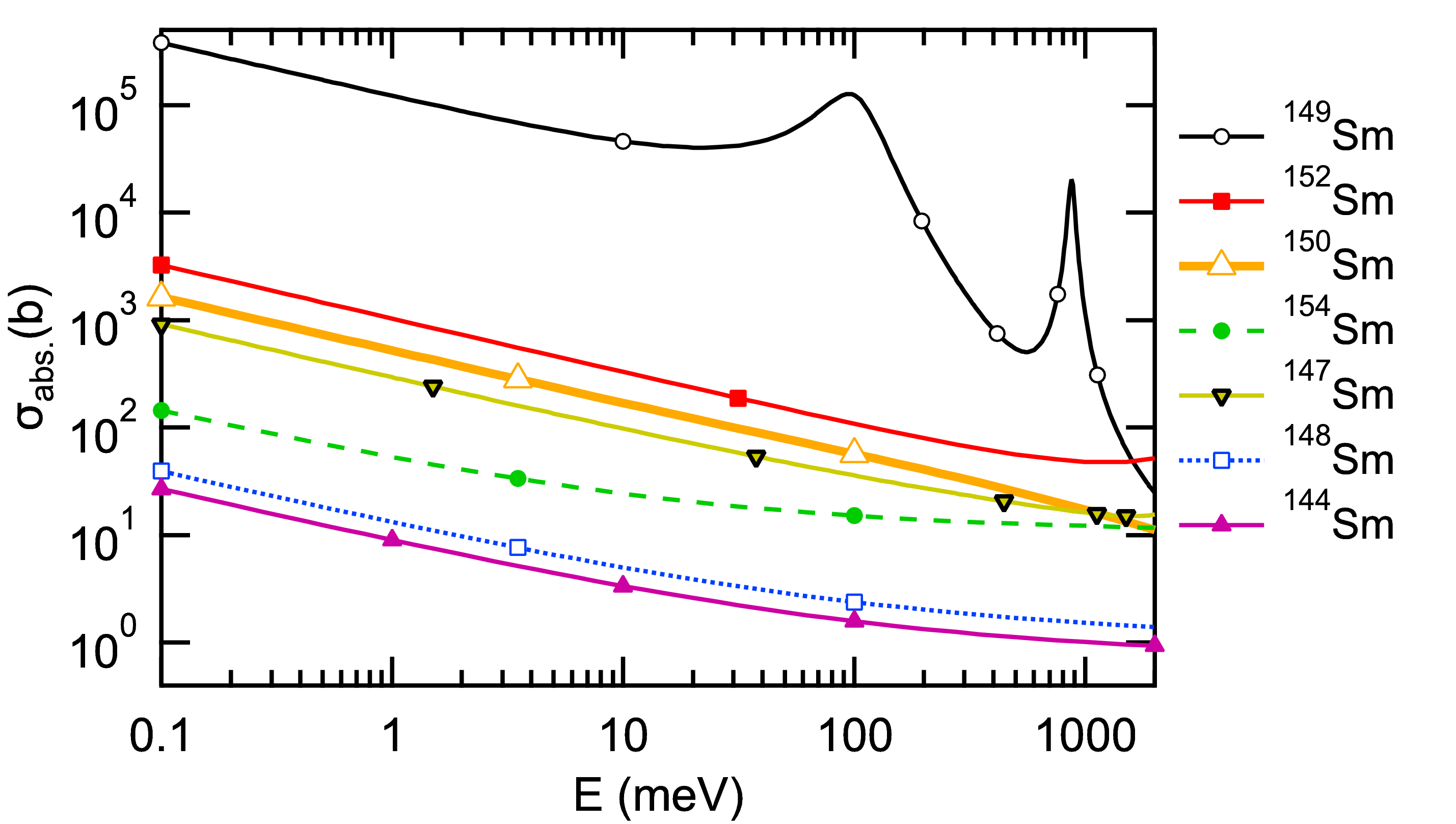}
\caption{
\label{fig: 0}
The neutron absorption cross section for different samarium isotopes are shown as a function of energy. The resonance near 100 meV for $^{149}$Sm is particularly important in this analysis.} 
\end{figure}

As described in the main paper, our nominal $^{154}$Sm$_2$Ti$_2$O$_7$ single crystal still contains enough neutron absorbing isotopes of Sm to significantly affect the details of the neutron spectroscopic measurements we performed. The neutron absorption correction is a property of the nucleus, and is thus unique to each isotope.  It is also a function of energy, and can display resonances in its energy spectrum.  Fig.~\ref{fig: 0} shows the energy dependence of the neutron absorption as a function  of energy for the isotope of Sm.  One can see that $^{149}$Sm is a particularly bad absorber, and it displays resonances in the absorption near 100 meV and 800 meV. The resonance near 100 meV is particularly relevant to the analysis of this experiment, as our neutron spectroscopy was performed with incident neutrons of $E_i=60$ meV and 150 meV. The former is very close to the 70.0(5) meV crystal field excitation measured at ARCS, and it could affect the intensity of this mode in a non-trivial way. We therefore carried out a Monte Carlo (MC) ray tracing analysis of full direct geometry time-of-flight chopper instruments using the MCViNE software package~\cite{lin2016mcvine}. This allows us to realistically estimate the effect of absorption, primarily from $^{149}$Sm. Our simulation has been performed using the instrument parameters of the ARCS experiment, which include an incident energy of $E_i = 150$ meV, a $T_0$ chopper frequency of 90 Hz, and a Fermi Chopper frequency of 600 Hz.

\begin{figure}[tbp]
\centering
\includegraphics[width=0.95\columnwidth]{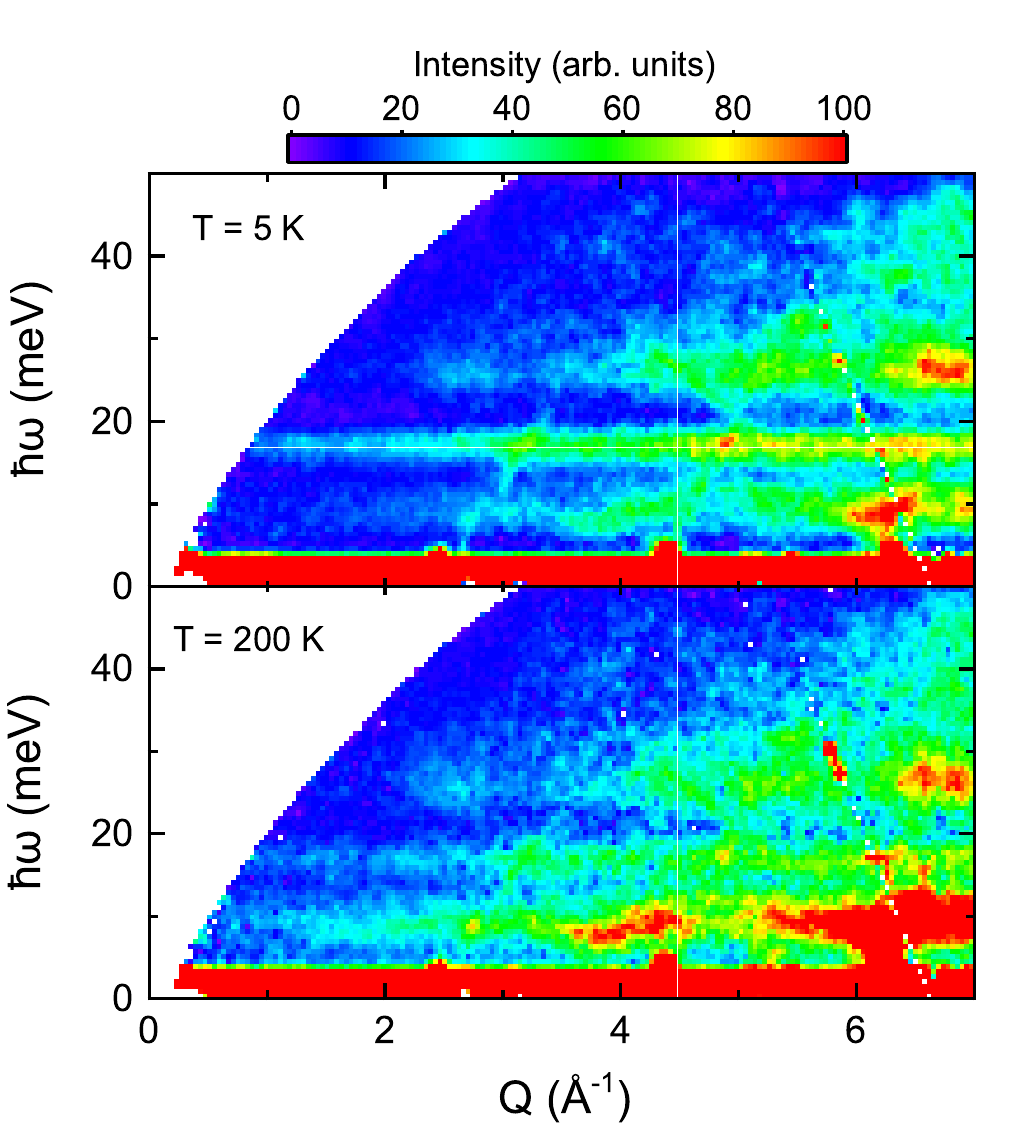}
\caption{
\label{fig: 0a}
$S(Q, \hbar \omega)$ for \STO~measured at ARCS with $E_i = 60$ meV. The data set clearly shows a crystal field transition at 16.3(5) meV. The other two excitations at 10 meV and 27 meV are phonon levels of Sm$^{3+}$ that have been observed in other rare earth pyrochlores (\emph{e.g.} Ref.~\cite{ruminy2016first}). Moreover, the unusual, non-monotonic form of the magnetic form factor for Sm$^{3+}$, wherein the intensity of this crystal field excitation decreases at low $Q$, is well captured by our Monte Carlo simulation.}
\end{figure}

Two simulations were performed to understand the absorption coefficient of \STO; one assuming zero absorption and one incorporating absorption effects by comparing the simulations to the neutron data shown in Fig.~\ref{fig: 0}.  The sample kernel used in the simulations was created assuming a cylindrical sample of \STO~with the same dimensions, absorption, isotopic content and orientation of the real sample. The crystal field transition observed in the ARCS experiment at 16.3(5) meV is shown in Fig.~\ref{fig: 0a}, the other two excitations at 10 and 27 meV are phonons in \STO~that have been observed in other rare earth pyrochlores (e.g. Ref.~\cite{ruminy2016first}). These crystal field transitions identified at 16.3(5) meV and 70.0(5) meV were simulated by introducing two non-dispersive levels in $Q$ with the correct magnetic form factor for Sm$^{3+}$. No sample environment or multiple scattering corrections were taken into account in this simulation. This calculation was performed in parallel on 30 cores, with 10 billion iterations to assure convergence and minimize the statistical noise.

\begin{figure}[htbp]
\centering
\includegraphics[width=0.95\columnwidth]
                {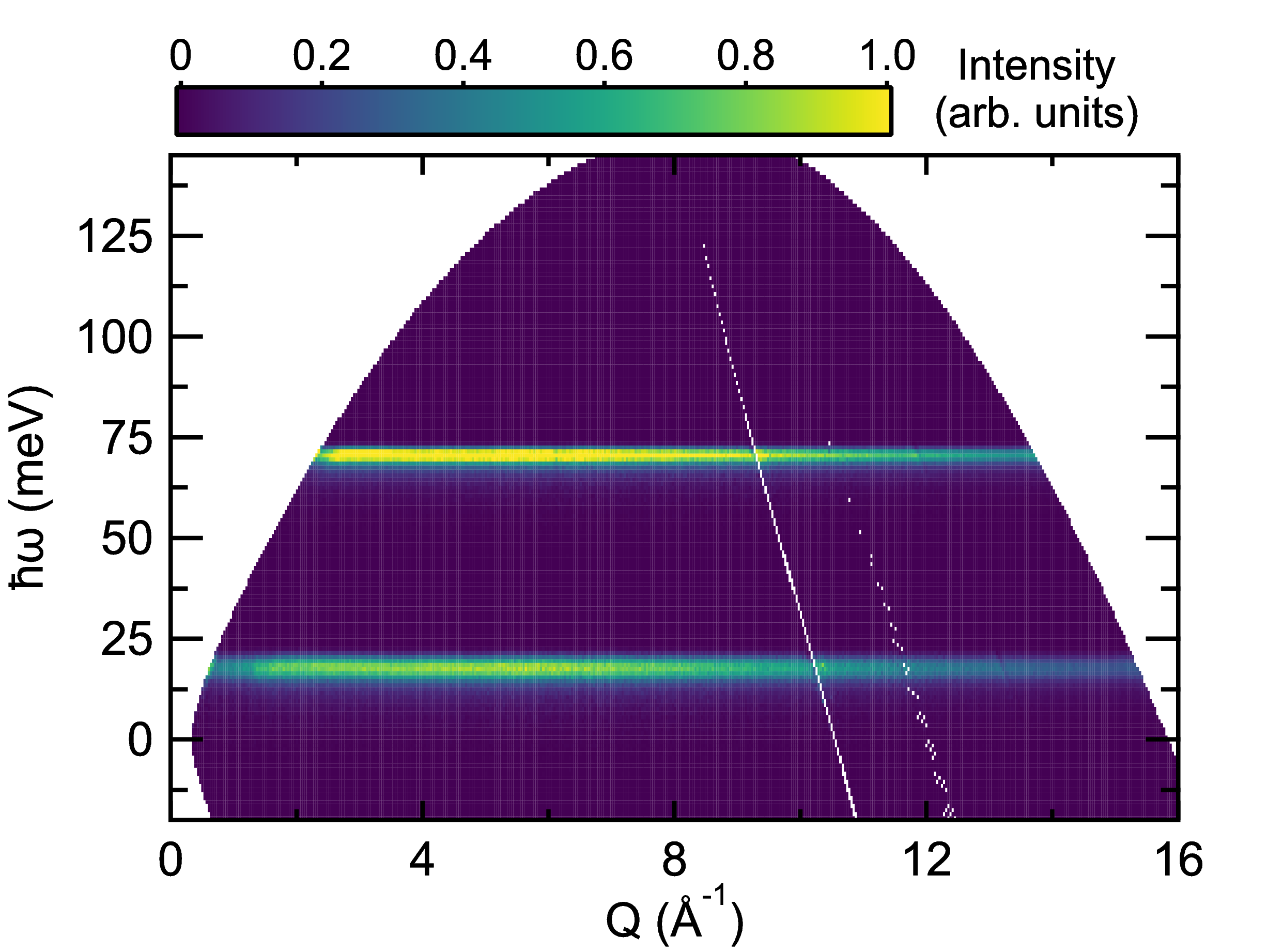}
\includegraphics[width=0.95\columnwidth]
                {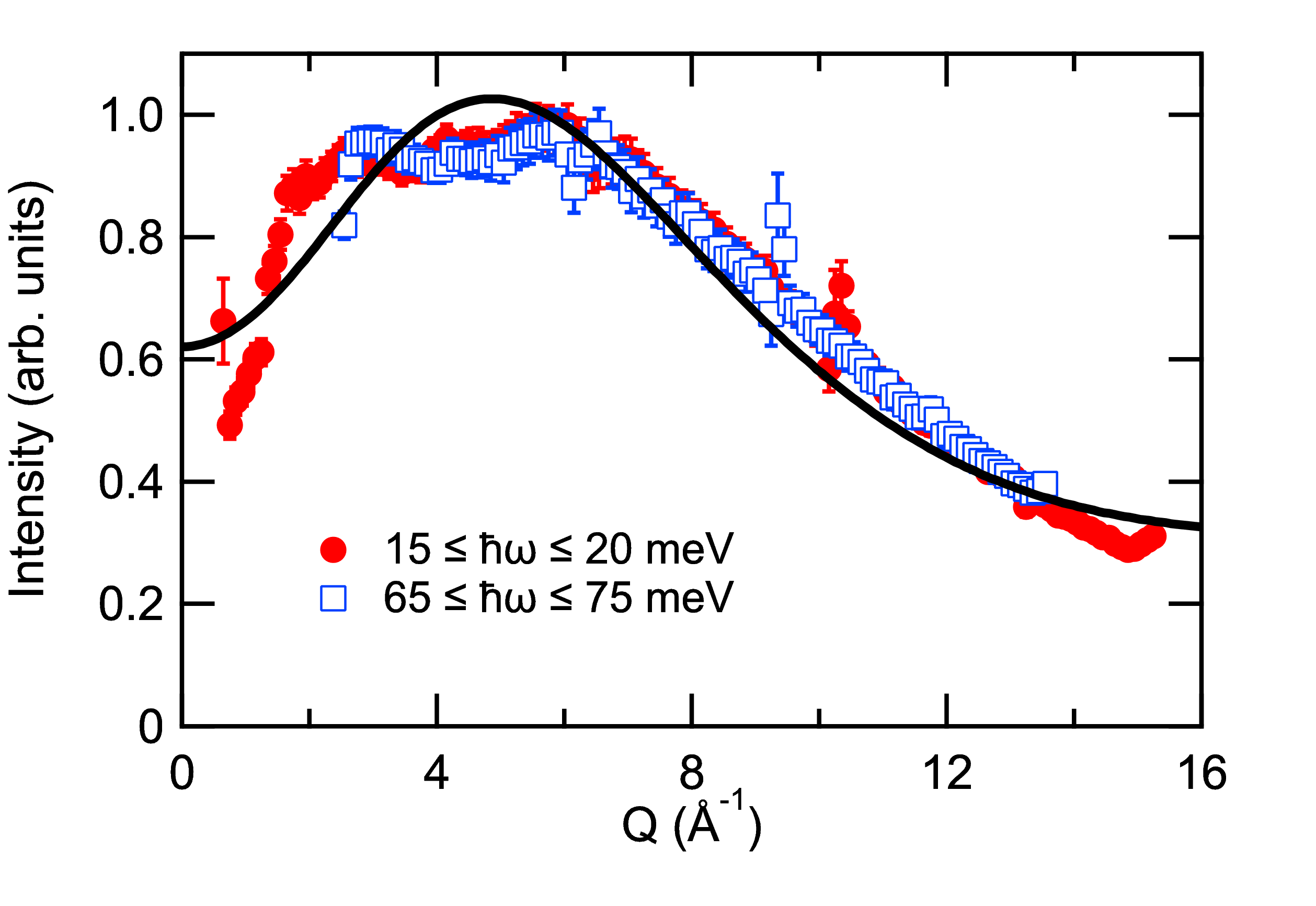}
\caption{
\label{fig: 1}
The result of the Monte Carlo simulation for \STO~are shown: \textbf{(a)} $S(Q,\omega)$ simulated for the  ARCS spectrometer with $E_i = 150$ meV and showing the two crystal field transitions at 16.3(5) and 70.0(5) meV. \textbf{(b)} Energy cuts across the two crystal field transitions are compared to the unusual magnetic form factor for Sm$^{3+}$, which is  non-monotonic as a function of $Q$.}
\end{figure}

\begin{figure}[htbp]
\centering
\includegraphics[width=0.95\columnwidth]
                {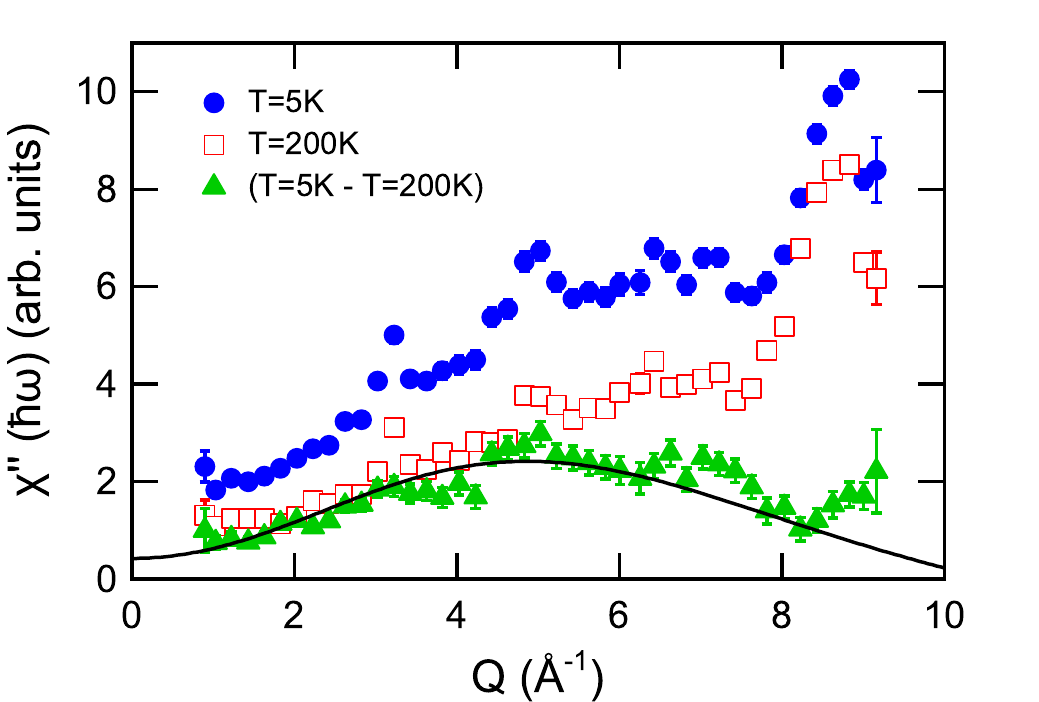}
\includegraphics[width=0.95\columnwidth]
                {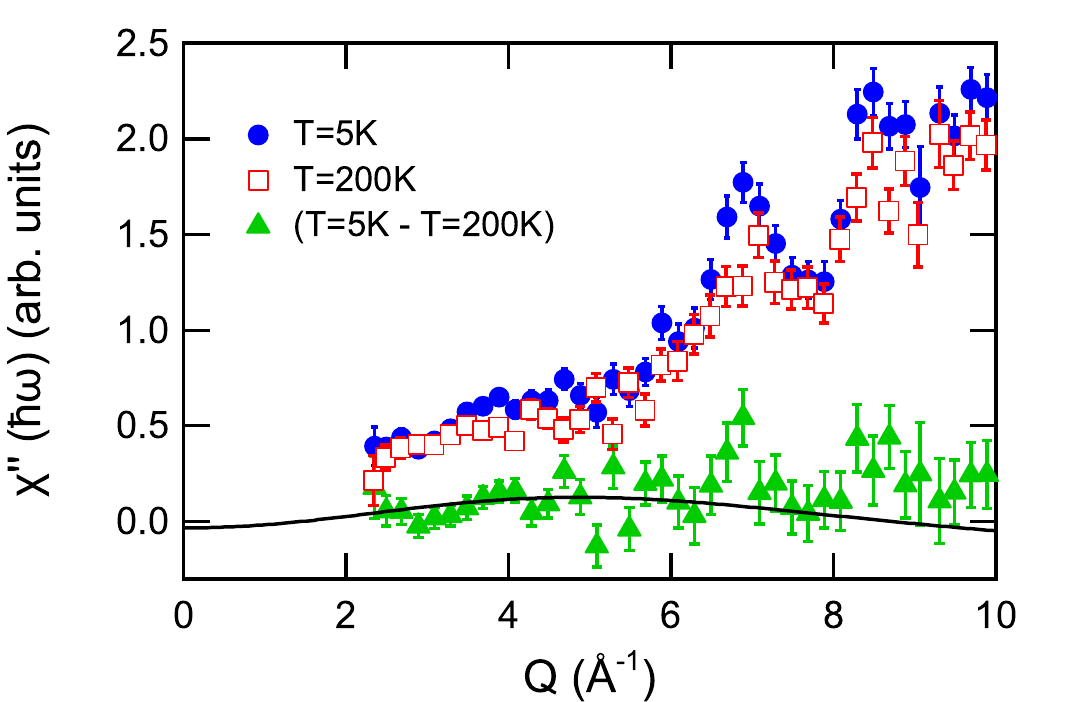}
\caption{
\label{fig: 2}
Energy cuts of the neutron data for \STO are shown. \textbf{(a)} A data set using $E_i = 60$ meV and showing the $Q$-dependence of the 16.3(5) meV  CEF transition is shown. \textbf{(b)} A data set using $E_i = 150$ meV and showing the $Q$-dependence of the 70.0(5) meV  CEF transition is shown.  The Q-dependence of the crystal field transitions measured at ARCS shows excellent agreement with the known Sm$^{3+}$ magnetic form factor. The fitting function consisted of a constant background, a small $Q^2$ term to account for a phonon background, and a multiplicative term times the square of the magnetic form factor. The Lande $g$-factor for Sm was fixed to 2/7.}
\end{figure}

The calculated $S(Q,\omega)$ of our MC simulation with an appropriate absorption correction is shown in Fig.~\ref{fig: 1}(a); individual energy cuts through the crystal field levels and showing their $Q$-dependence are compared to the known Sm$^{3+}$ magnetic form factor in Fig.~\ref{fig: 1}(b). The agreement between this MC simulation and the data set shown in Fig.~\ref{fig: 0}(a) is excellent. The absorption coefficient itself is estimated from the intensity ratio of the $Q$-cuts (integration range of 1 to 5~\AA$^{-1}$) for the two MC simulations, one with and one without absorption. This absorption coefficient was then applied to our data set before we proceeded with the crystal field analysis. As a consistency check, we also examine the $Q$-dependence of the crystal field transitions measured at ARCS in Fig.~\ref{fig: 2}. The blue and red markers represent the data collected at 5 K and 200 K respectively, while the green markers indicate the difference between the two data sets. The black line represents a fitting function consisting of a constant background, a small $Q^2$ term to account for a phonon background, and a multiplicative term times the square of the Sm$^{3+}$ magnetic form factor. The Lande $g$-factor for Sm was fixed to 2/7. The agreement between the data and the fit is excellent.

\subsection{\label{sec: 2}
Crystal Field Program 
        } 
        
In order to analyze the neutron scattering data and fit the crystal electric field (CEF) excitations we developed a calculation based on the point charge model~\cite{Hutchings1964227} and the Stevens' formalism~\cite{stevens1952matrix}. The former neglects the overlap between the orbitals and any relativistic corrections, while the latter is a mathematical tool to write an expansion of the Coulomb potential of the crystal based on the point group of the magnetic ion site. In our sample, 
the magnetic rare earth ion sits at the $A$-site of the pyrochlore lattice. Note that we rotated the reference system to align the local $\langle 111 \rangle$ direction along $\hat{z}$.

In general, the Coulomb potential of the crystal can be expressed using a linear combination of tesseral harmonics as follows,
\begin{eqnarray}
V(x, y, z) = \frac{q_j}{4 \pi \epsilon_{0}} \sum_{n=0}^\infty \frac{r^n}{R_{j}^{(n+1)}}\cdot Z 
\end{eqnarray}
\begin{eqnarray}
Z = \sum_{m} \frac{4 \pi}{(2n+1)} Z_{n m} (x_j , y_j , z_j) Z_{n m} (x, y, z).
\end{eqnarray}
Here $q_j$ is the charge of the ligand, $R_j$ is the position of the ligand and $Z_{n m} (x_j , y_j , z_j)$ is the tesseral harmonic~\cite{Hutchings1964227}. If we centre our reference system on the magnetic ion, we can rewrite the previous equation in this way,
\begin{eqnarray}
V(x,y,z) = \frac{1}{4 \pi \epsilon_{0}} \sum_{n}^\infty \sum_{m} r^n \gamma_{n m} Z_{n m} (x, y, z),
\end{eqnarray}
where for $k$ ligands,
\begin{eqnarray}
\gamma_{n m} = \sum_{j=1}^{k} \frac{q_j}{R_{j}^{(n+1)}}\frac{4 \pi}{2n +1} Z_{n m} (x_j , y_j , z_j).
\label{eq: 1.3} 
\end{eqnarray}
Equation~\ref{eq: 1.3} gives the coefficients of the linear combination of the tesseral harmonics. For every point group, only a few terms in the expansion are non-zero (see Ref.~\cite{walter1984treating}) and these terms coincide with the number of Stevens Operators we use in our Hamiltonian. The point group of both the scalenohedron and the trigonal anti-prism is $D_{3d}$ and thus, following Prather's convention~\cite{prather1961atomic}, only the terms $Z_{20}, Z_{40}, Z_{43}, Z_{60}, Z_{63}$ and $Z_{66}$ survive in our expansion. This convention states that the highest rotational $C_3$ axis of the system must form the $\hat{z}$ axis and that the $\hat{y}$ axis is defined as one of the $C_2$ axes. This assures that we have the minimum number of terms in the Coulomb expansion.

Finally we can use the so called ``Stevens Operators Equivalence Method" to evaluate the matrix elements of the crystalline potential between coupled wave functions specified by one particular value of the total angular momentum $J$. This method states that, if $f(x,y,z)$ is a Cartesian function of given degree, then to find the operator equivalent to such a term one replaces $x$, $y$, $z$ with $J_x$, $J_y$, $J_z$ respectively, keeping in mind the commutation rules between these operators. This is done by replacing products of $x$, $y$, $z$ by the appropriate combinations of $J_x$, $J_y$, $J_z$, divided by the total number of combinations. Note that, although it is conventional to use $J$ or $L$ in the equivalent operator method, all factors of $\hbar$ are dropped when evaluating the matrix elements.

As we are studying the ground state (GS) of a rare-earth system, without an external field applied, $S^2$, $L^2$, $J^2$ and $J_z$ are good quantum numbers. Thus the crystal field Hamiltonian can now be written as:
\begin{eqnarray}
H_{CEF} = const. \sum_{nm} \left[\frac{e^2}{4 \pi \epsilon_0} \gamma_{nm} \langle r^n \rangle \theta_n \right ] O_{n}^m \nonumber \\
= \sum_{nm} \underbrace{[A_{n}^m \langle r^n \rangle \theta_n  ]}_{B_{nm}} O_{n}^m = \sum_{nm} B_{nm} O_{n}^m,
\label{eq: 1.4}
\end{eqnarray}
where $\gamma_{nm}$ is the same coefficient as in Eq.~\ref{eq: 1.3}, $e$ is the electron charge, $\epsilon_0$ is the vacuum permittivity, $\langle r^n \rangle$ is the expectation value of the radial part of the wavefunction, $ \theta_n$ is a numerical factor that depends on the rare earth ion~\cite{Hutchings1964227}, $const.$ is a constant to normalize the tesseral harmonics and $O_{n}^m$ are 
the Stevens Operators.

The terms $A_{n}^m \langle r^n \rangle \theta_n$ are commonly called crystal field parameters, and they coincide with the parameters we fit in our calculation.
A general form of the Hamiltonian for our system is therefore:
\begin{eqnarray}
H_{CEF} = B_{20} O_{2}^0 + B_{40} O_{4}^0 +B_{43} O_{4}^3 \nonumber \\
+B_{60} O_{6}^0 +B_{63} O_{6}^3 +B_{66} O_{6}^6.
\label{eq: 1.5}
\end{eqnarray}      
Due to the fact that Sm$^{3+}$ has $L=5$, $S=5/2$ and $J = | L - S | = 5/2$, the $B_{6m}$ parameters in Eq.~\ref{eq: 1.5} are identically zero. The remaining three CEF parameters are then simultaneously varied to obtain the best agreement with the energies of the excitations and their relative intensity. The quantity that the calculation minimizes is:
\begin{eqnarray}
\chi^2 = \sum_i \frac{(\Gamma_{obs}^i - \Gamma_{calc}^i)^2}{\Gamma_{calc}^i},
\label{eq: 1.6}
\end{eqnarray}
where $\Gamma_{calc}$ is the calculated quantity of interest and $\Gamma_{obs}$ is the observed quantity.

Following this spirit, the logic of the calculation is the following:
\begin{enumerate}
\item Starting with an initial set of CEF parameters that can be calculated from first principles or taken from literature, we diagonalize our CEF Hamiltonian.
\item The eigenvalues are rescaled with respect to the ground state energy and we calculate and normalize the intensities of the CEF spectrum.
\item $\chi_{tot}^2 = \chi_{Energy}^2 + \chi_{Intensity}^2 + \chi_{Spectrum}^2  $ is calculated using Eq.~\ref{eq: 1.6}.
\item The procedure is iterated using another set of CEF parameters in order to minimize $\chi_{tot}^2$ until we converge on a solution which best estimates the experimental results.\
\end{enumerate}
The final CEF parameters $B_{nm}$ so obtained are then used to calculate the spectrum for a direct comparison with the data set. 

Examination of the Sm$^{3+}$ eigenvectors, presented in Table~1 of the main manuscript, show all three to be composed of a single set of time-reversed pairs of basis states. The first excited state at 16 meV is composed of pure $\ket{m_J=\pm \sfrac{1}{2}}$, while the second excited state at 70 meV is composed of pure $\ket{m_J=\pm \sfrac{5}{2}}$. A consequence of this is that the two excited state eigenvectors are not connected by dipole-allowed selection rules. Indeed, neutron scattering data collected at 200 K, where the 16 meV excited state would be thermally populated, shows no evidence for a thermally excited crystal field transition between the 16 and 70~meV levels, which would appear at approximately 54~meV. This is fully consistent with our assignment of the Sm$^{3+}$ eigenvectors and eigenvalues.

\subsection{\label{sec: 2}Scaling of Crystal Field Parameters} 

As described in the main text, we obtain a secondary confirmation of our crystal field solution using scaling arguments. This is achieved by starting from the CEF Hamiltonian for Er$_2$Ti$_2$O$_7$, which is highly constrained due to having a large number of ground state crystal field transitions~\cite{gaudet2017effect}. Following Ref.~\cite{Hutchings1964227} the scaling argument that connects the CEF parameters, $A_n^m(R)$, between pyrochlores with varying rare earths is:
\begin{equation}
A_n^m(R')=\frac{a^{n+1}(R)}{a^{n+1}(R')}A_n^m(R), 
\end{equation}
where $a(R) = 10.233$~\AA~is the cubic lattice parameter for Sm$_2$Ti$_2$O$_7$, taken from Ref.~\cite{knop1969pyrochlores}. This scaling procedure predicts a ground state doublet composed of pure $\ket{m_J = \pm \sfrac{3}{2}}$ well-separated from the first and second excited doublets, consistent with our CEF analysis.

\bibliography{Sm2Ti2O7_References}


